\documentclass[preprint]{aastex}

\usepackage{graphicx}
\usepackage{epsfig,latexsym}

\def\lapprox{\mathrel{\mathop  {\hbox{\lower0.5ex\hbox{$\sim$}
\kern-1.1em\lower-0.7ex\hbox{$<$}}}}}
\def\gapprox{\mathrel{\mathop  {\hbox{\lower0.5ex\hbox{$\sim$}
\kern-1.1em\lower-0.7ex\hbox{$>$}}}}}

\begin{document}

\title{The  chemical composition  of the Sun  from helioseismic
  and solar neutrino data} 

\author{Francesco L. Villante$^{1,2}$, Aldo M. Serenelli$^3$, Franck Delahaye$^4$ and Marc H. Pinsonneault$^5$}
\affil{$^1$Dipartimento di  Scienze Fisiche e Chimiche, Universit\`a  dell'Aquila, 67100 L'Aquila, Italy\\ 
$^2$Istituto Nazionale  di Fisica Nucleare (INFN), Laboratori  Nazionali del Gran
Sasso (LNGS), 67100 Assergi (AQ) , Italy\\
$^3$Instituto de Ciencias del Espacio (CSIC-IEEC), Facultad de Ciencias, 08193  Bellaterra, Spain\\
$^4$LERMA, Observatoire de Paris, ENS, UPMC, UCP, CNRS, 92190 Meudon, France\\
$^5$Astronomy Department, Ohio State University, Columbus, Ohio 43210, USA}

\begin{abstract}
We perform a quantitative analysis of the solar composition problem
by using a statistical approach that allows us to combine 
the information provided by helioseimic and solar neutrino data in
an effective way. We include in our analysis the helioseismic determinations of the
surface helium abundance and of the depth of the convective envelope,  
the measurements of the $^7{\rm Be}$ and $^8{\rm B}$ neutrino fluxes,  the 
sound speed profile inferred from helioseismic frequencies.
We provide all the ingredients to describe how  these  quantities depend  on  the  solar surface 
composition  and  to evaluate the (correlated) uncertainties in solar model predictions. 
We include errors sources that are not traditionally considered such as  those from inversion  of helioseismic  data. 
We, then, apply the proposed approach to infer the chemical composition of the Sun.
We show that the opacity profile of the Sun is well constrained by the solar observational properties.
In the context of a two parameter analysis in which elements are grouped as volatiles (i.e. C, N, O and Ne) 
and refractories (i.e Mg, Si, S, Fe), the optimal composition is found by
increasing the the abundance of volatiles by $\left( 45\pm 4\right)\%$ and that of refractories
by $\left( 19\pm 3\right)\%$ with respect to the values provided by
\cite{AGSS09}. This corresponds to the abundances $\varepsilon_{\rm O}=8.85\pm 0.01$ 
and $\varepsilon_{\rm Fe}=7.52\pm0.01$. 
As an additional result of our analysis, we show that
the observational data prefer values for the input parameters of the standard solar models 
(radiative opacities, gravitational settling rate, the astrophysical factors $S_{34}$ and $S_{17}$) that differ at 
the $\sim 1\sigma$ level from  those presently adopted. 
\end{abstract}


\maketitle

\newpage

%\begin{multicols}{2}

\section{Introduction}

In the last three decades, there has been enormous progress in our understanding of stellar structure and evolution.  
Solar models have played a particularly important role, in large part because we have powerful diagnostics of the internal solar 
conditions from solar neutrino experiments and helioseismology.  
The deficit of the observed solar neutrino  fluxes relative to solar model predictions, initially reported by Homestake~\citep{Homestake}
and then confirmed by GALLEX/GNO~\citep {Gallex,GNO}, SAGE~\citep{SAGE}, Kamiokande~\citep{Kamiokande} and Super-Kamiokande \citep{SK}, SNO~\citep{SNO} 
and Borexino~\citep{Borexino}, gave rise to the solar neutrino problem: major changes were required in either the theory of 
stellar structure and evolution and neutrino physics.  The development, refinement, and testing of the Standard Solar Model (SSM) played 
an important role in its ultimate resolution in 2002\footnote{For the sake of precision, the first model-independent evidence for solar neutrino oscillations 
and the first determination
of the $^8{\rm B}$ solar neutrino flux has been obtained in 2001 (see \citet{Fogli2001}) by comparing the SNO charged-current result \citep{SNOCC}
with the SK data with the method proposed by \citet{SKvsSNO}. The year 2002 is, however, recognized as the 'annus mirabilis' \citep{Fogli2003}
for the solar neutrino physics.}, 
when the SNO experiment obtained direct evidence for flavor oscillations 
of solar neutrinos and confirmed the SSM prediction of the $^8{\rm B}$ neutrino flux with a precision that, 
according to the latest data, is equal to about 3\%.

The Sun is a non-radial oscillator, and powerful insights have also emerged from the study of the solar frequency pattern (for example, see~\citet{BasuReview}).  
The sound speed as a function of depth can be reconstructed to high precision, of order 0.1\%.  
Abrupt changes in the solar thermal structure from ionization and the transition from radiative to convective energy transport induce acoustic glitches 
that can be precisely localized; we can therefore infer the depth of the convective envelope at the 0.2\% level and the surface helium abundance at the 1.5\% level. 
As a  result, the solar structure is now well constrained and the Sun can be used as a solid benchmark
for  stellar evolution  and  as a  \emph{laboratory}  for fundamental  physics
(see e.g. \citet{fiorentini:2001,ricci:2002,bottino:2002}).  In fact, it was the excellent
agreement between solar models 
and helioseismic  inferences on the solar structure  (better than 1.5$\sigma$
for all constraints) that  gave a strong support to the idea  that the root of
the \emph{solar neutrino  problem} had to be found outside  the realm of solar
modelling \citep{BP01}, before evidence for neutrino oscillations was found.

All these important measurements  acquire even more relevance when considering
that,  in recent  years, a  \emph{solar  composition  problem}  has emerged.  
The SSM treats the absolute and relative elemental abundances as an input, and the 
\citet{GS98} (hereafter GS98) mixture yields concordance between model and data.  
Relative abundances of heavy elements can be precisely measured in meteorites \citep{Lodders}, 
but the abundances of the important light CNO elements can only be measured in the photosphere.  
The ${\rm Ne}$ abundance is even less secure, as it is inferred from solar wind measurements.  
A systematic overhaul in solar model atmospheres, see \citet{AGS05} and \citet{AGSS09} (hereafter AGSS09), 
has led to a downward revision in the inferred photospheric heavy element abundances (see Table~\ref{Tab1}) by up to 30-40\% 
for important species such as oxygen.  
The magnitude of the differences is model(er) dependent; independent measurements by \citet{Caffau} 
(see also \citet{Lodders})  are intermediate between the GS98 and AGSS09 scales.  The internal structure of SSMs 
using the lower solar surface metallicity of AGSS09 does not reproduce the helioseismic constraints;  for example, the sound speed disagrees at the bottom
of the convective envelope by about $\sim 1\%$ with the value inferred from helioseismology. In addition,
the predicted surface helium abundance is lower by $\sim7\%$ and the radius of the convective envelope
is larger by $\sim1.5\%$ with respect to the helioseismic results. In synthesis, inferences from modern three-dimensional hydrodynamic models of the solar atmosphere lead to predictions
for the solar interior that are in strong disagreement with observational constraints, well above the
currently estimated errors.  

The solar compositon problem has been addressed by a number of independent interior calculations. 
Problems in reconciling helioseismic data with a low abundance scale became obvious early on (\citet{BasuYs}; \citet{BP2004prl}; \citet{Turck2004}).
 \citet{Bahcall06} performed a Monte Carlo analysis that included calculations of the RMS deviations between the inferred
solar sound speed profile and that predicted by models with the "high" and "low" abundances. \citet{pinso} advocated 
an inversion of the problem, solving for the abundances consistent with the base of the surface convection zone and surface helium abundance.  
The ionization signature in the surface convection zone can also be used as an independent diagnostic suggesting high metallicity 
(see, however, discussion in  \citet{Vorontsov}); \citet{BasuReview} summarized the initial results favoring a higher solar metallicity.  
Physical processes not included in the standard model could potentially provide an explanation, but the combination of constraints has proven difficult
 to reproduce in practice \citep{Guzik}.  An intermediate solar metallicity using low-degree modes was reported 
by \citet{Houdek}; however, these authors note that their mixture produces a sound speed profile at variance with solar data. 
%Systematic differences in the acoustic depth derived from low order modes relative to prior ones using the more robust higher order modes available in the Sun are the probable explanation. 

The  goal of this  work is  to perform a complete and quantitative  analysis of  the solar
composition problem. In particular,  we address the following questions: 
which is the chemical composition of the  Sun that, by using the current input physics of solar models, 
can be inferred from helioseismic and solar neutrino data? 
How does different observational information combine
in  determining the  optimal composition  of the  Sun? How  does  the obtained
composition  compare to  the photospheric  inferred values?  Do  the different
observational data show tensions and/or inconsistencies that may point at some
inadequacies in the SSM input parameters or assumptions?  Even if the problem
has  been  already  considered   in  literature,  a  thorough  self-consistent
discussion is still missing. While a  rigorous approach is not necessary for a
qualitative  assessment of  the problem,  it becomes  essential for  our goal,
i.e.  to  use the  helioseismic  information  in  combination with  the  solar
neutrino  results to  infer the  properties of  the Sun.  In order  to  make a
correct inference,  one has to  define an appropriate figure-of-merit  (e.g. a
$\chi^2$ statistics)  that has  to be non-biased  and should  combine the
different pieces  of the observational  information with the  correct relative
weights.

In this respect, important progress has  to be done at a methodological level.
This papers starts  addressing this problem.  We propose  to use a statistical
approach,  normally  adopted in  other  areas  of  physics (e.g.  in  neutrino
studies, see \cite{lisi}) in which  all the relevant pieces of information can
be combined in a correct and  effective way.  We discuss a strategy to include
the observational information  for the sound speed profile,  the radius of the
convective envelope,  the surface helium  abundance, and the $^7{\rm  Be}$ and
the $^8{\rm  B}$ neutrino fluxes. We  provide all the  ingredients to describe
how  these  quantities depend  on  the  assumed  chemical composition  and  to
evaluate the (correlated) uncertainties in solar model predictions.

The plan  of the paper is  as follows.  In \S~\ref{sec:modsdata},  we review the
status of SSM  calculations, discuss in detail our  treatment of uncertainties
in theoretical  predictions and also the  observational constraints considered
in the analysis, including sources of errors not traditionally considered such
as  those from inversion  of helioseismic  data.  In  \S~\ref{sec:stat}, we
describe  the  adopted statistical  approach.   In \S~\ref{sec:metals},  we
analyze  the response of  the Sun  to variations  of its  surface composition.
\S~\ref{sec:results} contains the results of our analysis, i.e.  the bounds
on the chemical composition of the Sun that are inferred from helioseismic and
solar  neutrino  data.  Finally,  we  provide  a  summary and  conclusions  in
\S~\ref{sec:conclusions}.

\section{Models and Data}
\label{sec:modsdata}

Our  theoretical working  framework  is  the SSM  and  in \S~\ref{sec:ssm}  we
summarize the aspects most relevant to this work. More importantly, we present
our  treatment of  errors,  for which  two qualitatively  different
sources must be  identified.  On one hand, errors in the input parameters $I$
for solar model construction induce \emph{theoretical} uncertainties in the 
SSM predictions $Q$. These errors are fully correlated, 
as it is discussed in \S~\ref{sec:theo}.
On  the other, the observational determinations $Q_{\rm obs}$ of helioseismic quantities 
and solar neutrino fluxes are affected by {\em observational} errors. In this work, 
we treat these errors as  uncorrelated, as it is discussed in \S~\ref{ObsConstraints}.

\subsection{Standard Solar Models}
\label{sec:ssm}

%-----------------------------------------------------

\begin{table}[t]
\begin{center}{
\small
\begin{tabular}{c|cc| c}
\hline \hline
 Element & AGSS09 & GS98 & $\delta  z_i$\\ \hline C & $8.43 \pm 0.05$ &
 $8.52\pm 0.06$ &  0.23 \\ N & $7.83  \pm 0.05$ & $7.92\pm 0.06$ &  0.23\\ O &
 $8.69\pm  0.05$ &  $8.83\pm  0.06$ &  0.38\\  \hline Ne  &  $7.93\pm 0.10$  &
 $8.08\pm  0.06$ &  0.41\\ \hline  Mg &  $7.53\pm 0.01$  & $7.58  \pm  0.01$ &
 0.12\\ Si  & $7.51\pm 0.01$ &  $7.56\pm 0.01$ &  0.12\\ S & $7.15\pm  0.02$ &
 $7.20\pm 0.06$ & 0.12\\ Fe &  $7.45\pm 0.01$ & $7.50\pm 0.01$ & 0.12\\ \hline
 \hline $Z/X$ & 0.0178 & 0.0229 & 0.29 \\ \hline
\end{tabular}
}\end{center}
%\vspace{0.4cm} 
\caption{{\protect
\label{Tab1} 
Solar surface heavy element abundances in the AGSS09 \citep{AGSS09} and GS98 
\citep{GS98} admixtures.   Abundances are given  as $\varepsilon_{j}\equiv \log
\left(N_j/N_H\right)+12$, where $N_j$ is the 
number density of element $j$. Last row gives the total metal-to-hydrogen
mass fraction. 
In the last column, we  show the fractional differences $\delta z_{j}$ between
the abundances in the two compilations.}}
\vspace{0.4cm}
\end{table}

%---------------------------------------------------
The development of a new generation of stellar atmospheres models resulted in a downward 
revision of the solar photospheric abundances
%The most recent determinations of the solar photospheric abundances, which are
%based  on well-tested sophisticated  three-dimensional hydrodynamic  models of
%the  solar atmosphere, indicate  that the  solar metallicity  is substantially
%lower  with  respect  to  previous  estimates.   
This  is  shown  e.g.   in
Table \ref{Tab1} where we give the abundances of C, N, O, Ne, Mg, Si, S and Fe,
which are  the relevant  elements for solar  model construction for  the (new)
AGSS09 \citep{AGSS09} and (old) GS98 \citep{GS98} heavy element admixtures.
Last  column  shows  the  fractional  differences   between  the  individual
abundances (relative to hydrogen) in the two compilations. 

The heavy element  admixture determines to a large  extent the opacity profile
of the Sun and is a crucial  input for solar model construction.  This is seen
in the first  two columns of Table~\ref{Tab2}, where we  report the results of
two  recent SSM calculations~\citep{serenelli:2011}  that implement  the AGSS09
\citep{AGSS09} and GS98 \citep{GS98}  surface compositions. The models have been
computed with  GARSTEC \citep{garstec:2008},  using the nuclear  reaction rates
recommended   in   \citet{sfii} and    the   input   physics   described   in
\citet{serenelli:2009}.  Element diffusion in the solar interior is included according
to \citet{thoul}. 
%The updated 2006  OPAL equation of state is used \citep{Rogers1,Rogers2}.
The models have been computed by using the  radiative opacities from  the Opacity
Project  (OP; \citet{OP}),  complemented  at  low  temperatures with  the  opacities  from
\cite{ferguson2005}.   We consider  the following  observable  quantities: the
surface helium abundance $Y_{\rm b}$, the  radius of the inner boundary of the
convective  envelope $R_{\rm b}$,  the neutrino  fluxes $\Phi_\nu$,  where the
index  $\nu =  {\rm  pp, Be,  B,  N, O}$ refers  to  the neutrino  producing
reactions   according   to   the   usual   convention.   We   also   show   in
Figure~\ref{Fig1} the fractional difference $ \delta c_{i}\equiv \left( c_{\rm
  obs, i}- c(r_{i}) \right)/c(r_{i})$ between the predicted sound speed $c(r)$
and the values $c_{\rm obs,i}$ inferred from helioseismic data; the black line
refers the SSM model implementing the AGSS09 surface composition while the red
line is obtained by using the GS98 admixture.

In the  following, we take  the SSM implementing  the AGSS09 composition  as a
starting  point  for our  analysis  and  we  use the  notation  $\overline{Q}$
($\overline{I}$)  to  indicate the  prediction  (assumption)  for the  generic
quantity $Q$ (input $I$) in this calculation.

\begin{table}[t]
\begin{center}{
\small
\begin{tabular}{c|cc|c|c}
\hline
\hline
 & AGSS09 & GS98 & Obs. & GS98rec \\ \hline
$Y_{\rm b}$  & $0.2319\,(1\pm 0.013)$  & $0.2429\,(1\pm 0.013)$ &  $0.2485 \pm
0.0035 $ & 0.243 \\ 
$R_{\rm  b}/R_\odot$ &  $0.7231\,(1\pm  0.0033)$ &  $0.7124\,(1\pm 0.0033)$  &
$0.713 \pm 0.001 $ & 0.710 \\ 
\hline
$\Phi_{\rm    pp}$    &    $6.03\,(1\pm0.005)$   &    $5.98\,(1\pm0.005)$    &
$6.05(1^{+0.003}_{-0.011})$ & 5.98\\ 
$\Phi_{\rm    Be}$   &    $4.56\,(1\pm0.06)$   &    $5.00    \,(1\pm0.06)$   &
$4.82(1^{+0.05}_{-0.04})$ & 4.98 \\ 
$\Phi_{\rm B}$ & $4.59\,(1\pm  0.11)$ & $5.58\,(1\pm0.11)$ & $5.00(1\pm 0.03)$
& 5.49 \\ 
$\Phi_{\rm N}$ & $2.17\,(1\pm0.08)$ & $2.96\,(1\pm0.08)$ &$\le 6.7$ & 2.89\\
$\Phi_{\rm O}$ &$1.56\,(1\pm0.10)$ & $2.23\,(1\pm0.10)$ & $\le 3.2$ & 2.15\\
\hline
\end{tabular}
}\end{center}
\vspace{-0.5cm} 
\caption{ {\protect The   predictions  of  SSMs  implementing  GS98
    \citep{GS98}   and  AGSS09   \citep{AGSS09}  admixtures.    The  theoretical
    uncertainties have been  calculated as it is described in  the text and do
    not include  the contributions due  to errors in the  surface composition.
    In the  third column,  we show the  observational values  for helioseismic
    quantities     \citep{BasuYs,BasuRb}    and    solar     neutrino    fluxes
    \citep{BorexinoPC}.  In the last column,  we calculate the GS98 solar model
    predictions starting from AGSS09 solar model by using the linear expansion
    given  in  equation (\ref{linear}).   The  neutrino  fluxes  are given  in
    following units: $10^{10}\,{\rm  cm}^{-2}{\rm s}^{-1}$ (pp); $10^{9}\,{\rm
      cm}^{-2}{\rm  s}^{-1}$ (Be);  $10^{6}\,{\rm  cm}^{-2}{\rm s}^{-1}$  (B);
    $10^{8}\,{\rm   cm}^{-2}{\rm  s}^{-1}$  (N);   $10^{8}\,{\rm  cm}^{-2}{\rm
      s}^{-1}$ (O).
\label{Tab2} 
}}\vspace{0.5cm}
\end{table}

\subsection{Theoretical uncertainties}
\label{sec:theo}

\begin{figure}[t]
\par
\begin{center}
\includegraphics[width=12cm,angle=0]{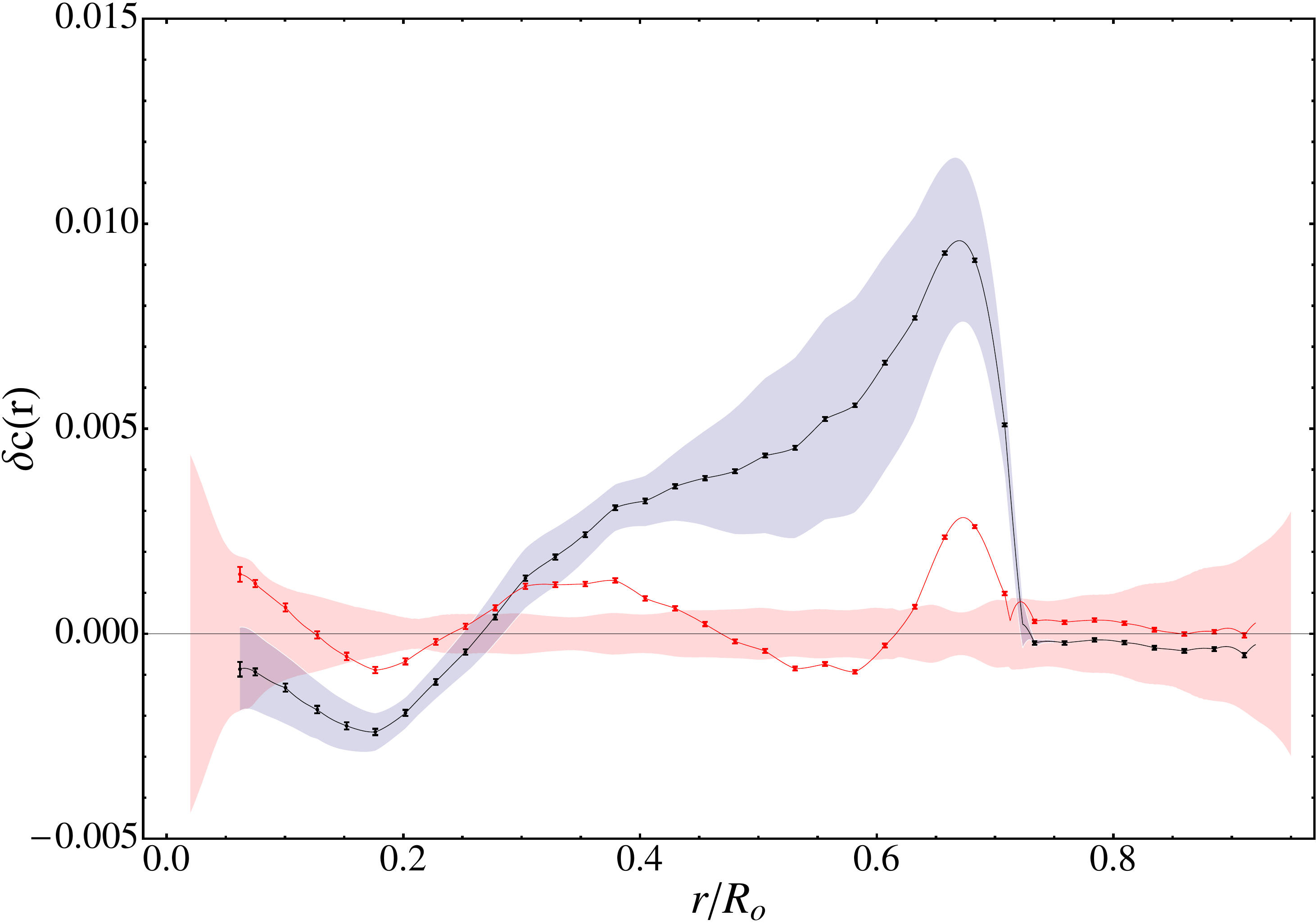}
\end{center}
\par
%\vspace{-5mm} 
\vspace{-0.5cm}
\caption{\protect The  fractional  difference $  \delta  c_{i}\equiv
  \left( c_{\rm  obs, i}- c(r_{i})  \right)/c(r_{i})$ between the  sound speed
  $c(r)$  predicted by  SSMs  and  the values  $c_{\rm  obs,i}$ inferred  from
  helioseismic data; the  black line refers to the  SSM model implementing the
  AGSS09 surface composition while the red  line is obtained by using the GS98
  admixture. The red  band provides an estimate of  the uncertainty
  in  inversion  of helioseismic  data.   The  light blue  band corresponds  to  the
  $1\sigma$ uncertainties in the theoretical predictions.}
\vspace{0.5cm} 
\label{Fig1}
\end{figure}

\begin{figure*}[t]
\par
\begin{center}
\includegraphics[width=8.cm,angle=0]{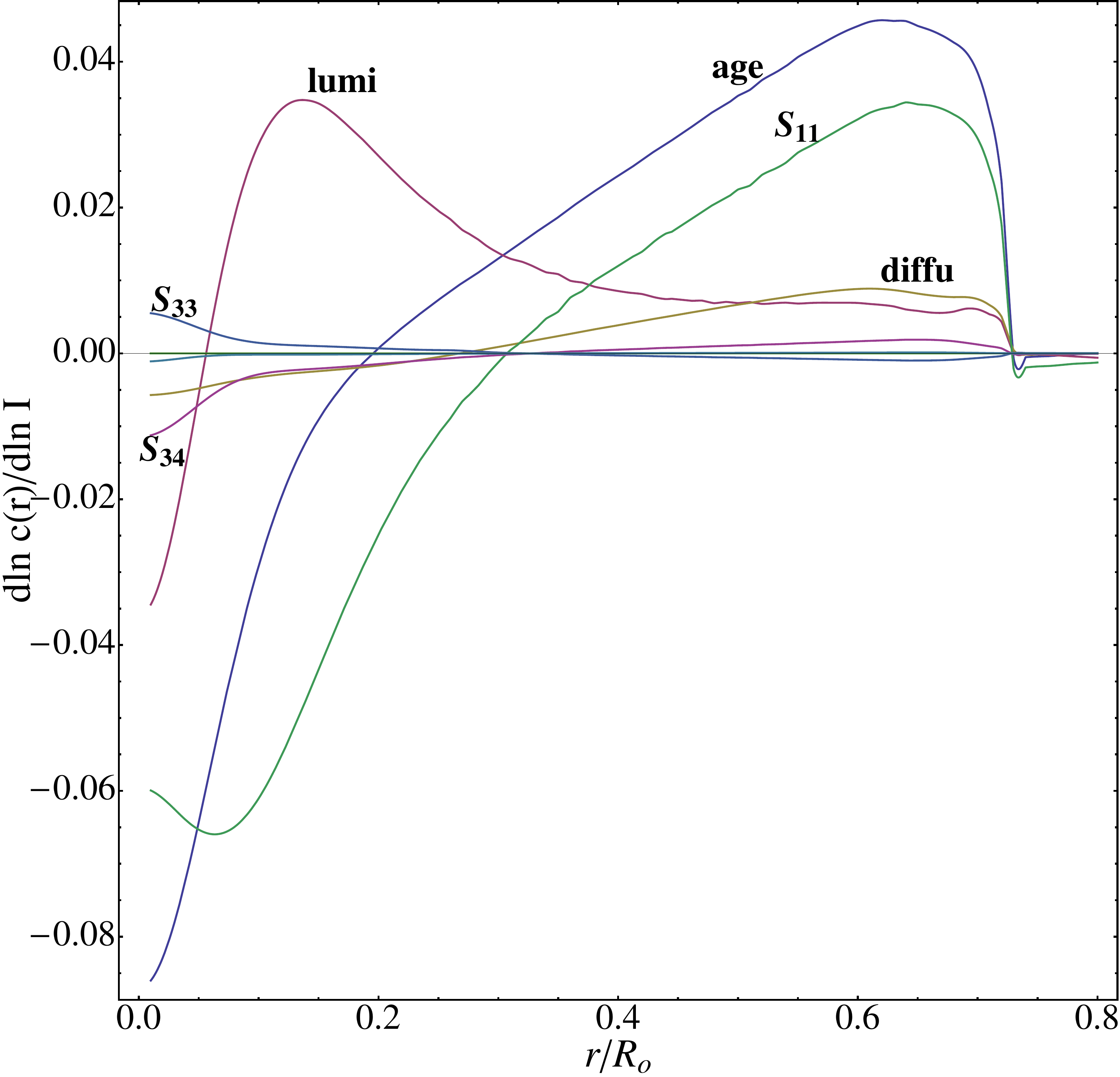}
\includegraphics[width=8.cm,angle=0]{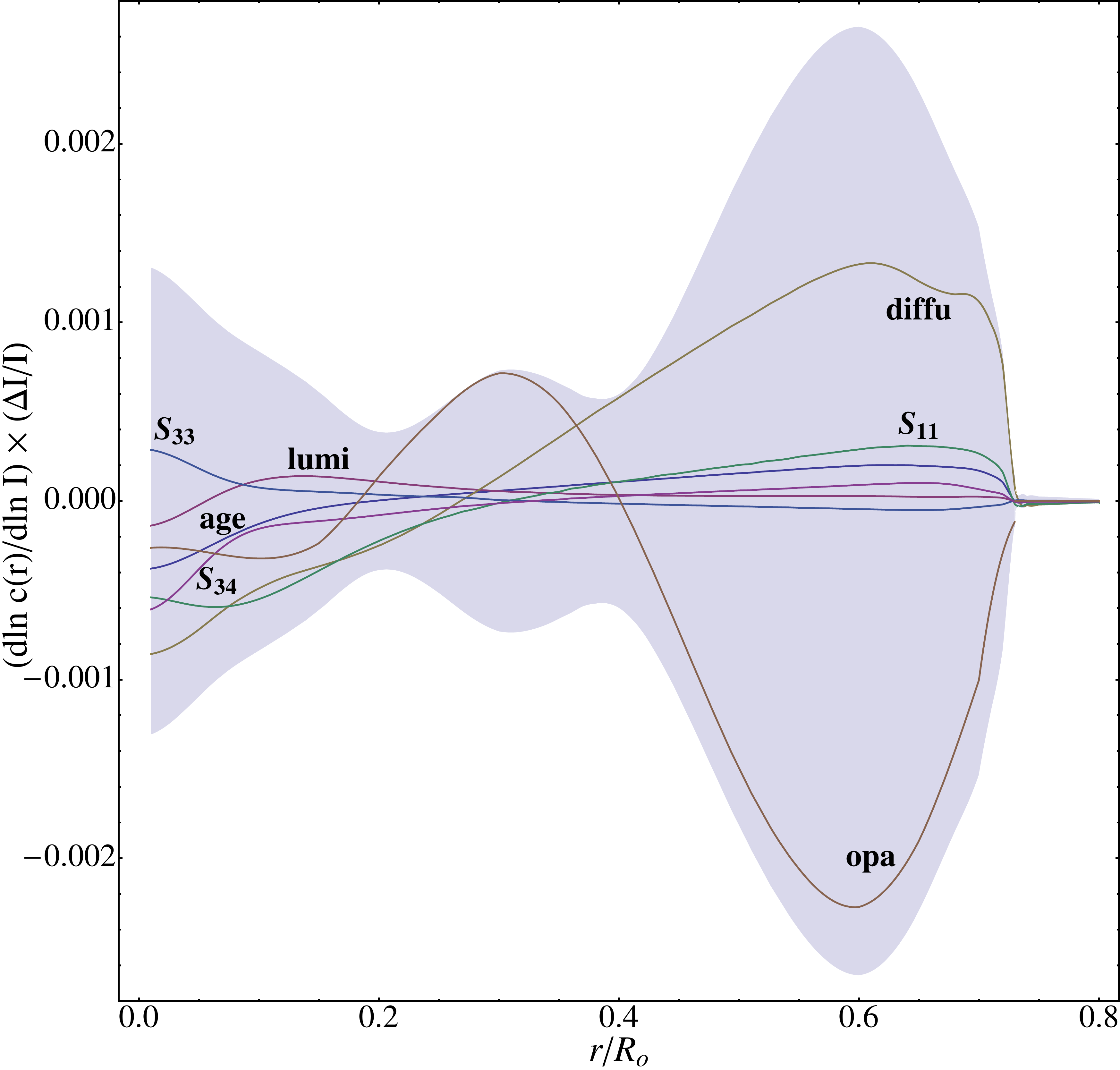}
\end{center}
\par
\vspace{-0.5cm} 
\caption{\protect Left panel:  The logarithmic  derivatives of
  the  sound speed  profile defined  in equation~(\ref{SoudSpeedDer}).  Right
    panel: The  contributions to uncertainties in  the theoretical prediction
  of the sound  speed profile defined in eqs.(\ref{TheoErr},\ref{TheoErrOPA}).
  The shaded area corresponds to the total theoretical error.}
\vspace{0.5cm}
\label{Fig2}
\end{figure*}

The   errors  quoted   in  Table~\ref{Tab2}   and  the   light-blue   band  in
Figure~\ref{Fig1}  correspond   to  $1\sigma$  uncertainties   in  theoretical
predictions.   They have  been calculated  by  propagating the  errors in  the
following input parameters:  the age of the Sun  (\texttt{age}); the diffusion
coefficients  (\texttt{diffu});  the  luminosity (\texttt{lum});  the  opacity
profile  (\texttt{opa})  of  the  Sun;  the  astrophysical  factors  $S_{11}$,
$S_{33}$, $S_{34}$,  $S_{17}$, $S_{e7}$ and  $S_{1,14}$. Note that, as  we are
interested in establishing  bounds on the solar chemical  composition, we have
explicitly  omitted   its  contribution  to   theoretical  uncertainties.   By
following  the   standard  procedure,  we  have   calculated  the  logarithmic
derivatives
\begin{equation}
B_{Q, I} = \frac{d \ln Q}{d \ln I}. 
\end{equation}
$B_{Q,I}$ values are   available   at \cite{serenelli:2013} for  all  observables  except  the  sound  speed
$c(r)$. To our knowledge, the  logarithmic derivatives $B_{c,I}(r)$ of the  sound speed, defined
as
\begin{equation}
B_{c, I}(r) \equiv d\ln c(r)/ d\ln I
\label{SoudSpeedDer}
\end{equation}
have not  been given elsewhere in  the scientific literature and  are shown in
the left panel of Figure~\ref{Fig2} as a function of the solar radius.

As mentioned before, the uncertainty $\delta I$ in each input parameter $I$ produces fully 
correlated errors on the SSM predictions of observables $Q$. To emphasize this  
point, we use the  symbol $C_{Q,  I}$ to  indicate the  fractional variation  
of $Q$  when a fractional correction $\delta  I$ is applied to the  input $I$.  
The  various contributions $C_{Q,I}$, shown in Table~\ref{Tab4} and in 
the right panel of Figure~\ref{Fig2}, are calculated from the relation
\begin{equation}
C_{Q, I} = B_{Q, I}  \; \delta I 
\label{TheoErr}
\end{equation}
with the $\delta I$ values summarized in \cite{serenelli:2013}. The only exception 
is the opacity, that we discuss below.
%in tab.~\ref{Tab3}.

%=================================================================

%The  opacity contributions  $C_{Q,\rm opa}$  to theoretical  uncertainties are
%also  obtained  through a  linearized  approach, as  it  is  discussed in  the
%following. 
Opacity  is not a single  number but a complicated  function of the
properties of the solar plasma which can  be modified in a non trivial way. To
take this into account, we use the opacity kernels derived in \cite{villante2}
by  adopting the  linearization procedure  proposed in  \cite{villante1}.  The
kernels $K_{Q}(r)$ represent the  functional derivatives of the observable $Q$
with respect to  the opacity profile of  the Sun and can be  used to calculate
the  effects produced  by an  arbitrary opacity  variation  $\delta \kappa(r)$
according to:
\begin{equation}
\delta Q = \int dr \; K_{Q}(r) \; \delta \kappa(r)
\label{KappaEff}
\end{equation}
where:
\begin{equation}
\delta Q \equiv \frac{Q}{\overline{Q}}-1 \;,
\end{equation}
and
\begin{equation}
\label{kappa_int}
\delta                             \kappa(r)                            \equiv
\frac{\kappa(\overline{T}(r),\overline{\rho}(r),\overline{Y}(r), 
\overline{Z}_j(r))}{\overline{\kappa}(\overline{T}(r),\overline{\rho}(r), 
\overline{Y}(r),\overline{Z}_j(r))} - 1 
\end{equation}
In  the above  relation, the  functions $\kappa$  and  $\overline{\kappa}$ are
calculated along  the density,  temperature and chemical  composition profiles
predicted by SSM.

%=================================================================

%===============================================

\begin{figure}[t]
\par
\begin{center}
\includegraphics[width=10cm,angle=0]{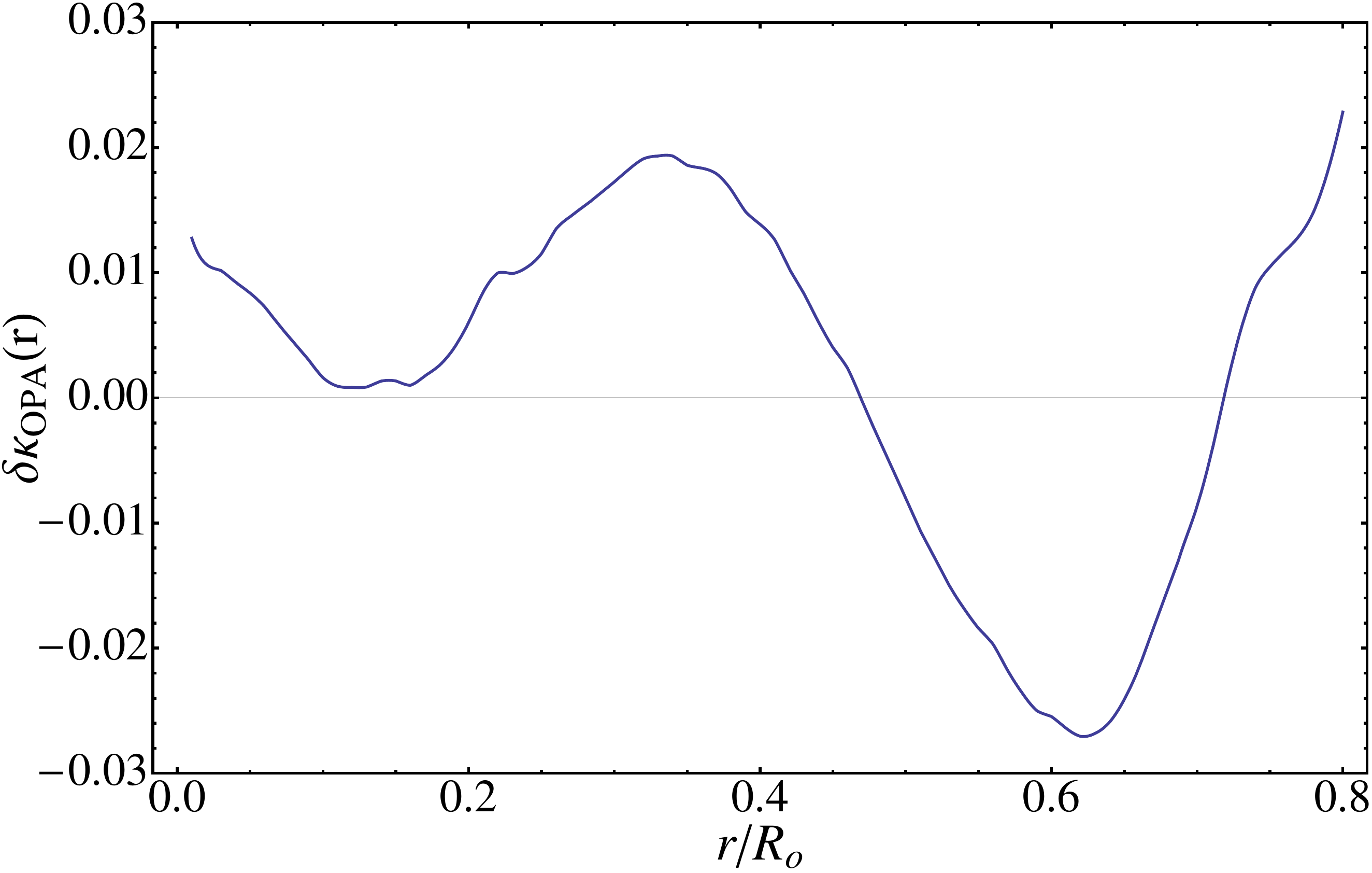}
\end{center}
\par
\vspace{-0.5cm} 
\caption{\protect \label{Fig3} Fractional difference between OPAL
  and  OP   radiative  opacities  calculated   along  the  SSM   profile,  see
  equation~(\ref{dkappaOPALOP}).}
\vspace{0.5cm}
\end{figure}

%----------------------
By using the kernels $K_{Q}(r)$, we propagate the 
uncertainty in the opacity profile of the Sun $\delta \kappa_{\rm opa}(r)$ 
according to:
\begin{equation}
C_{Q, \rm opa} \equiv \int dr \; K_{Q}(r) \; \delta \kappa_{\rm opa}(r)
\label{TheoErrOPA}
\end{equation}
The  standard  prescription, adopted  e.g.   in  \cite{serenelli:2009}, is  to
consider a  $2.5\%$ global rescaling  factor that corresponds to  take $\delta
\kappa_{\rm opa}(r)  \equiv 0.025$.  However,  this is not  realistic because,
albeit different groups typically provide  opacity values that differ by $\sim
{\rm few}\%$ in  the solar interior, there is a  complicated dependence on the
solar radius. Moreover,  it was shown in \cite{villante2}  that the assumption
of a global rescaling underestimates the uncertainties for the sound speed and
for  the depth  of  the convective  envelope.  The opacity  kernels for  these
quantities are not positive definite  
and a constant $\delta \kappa$ produces effects in different regions of the Sun 
that partially compensate each other.
For this reason,  we take the difference between the
OP \citep{OP} and OPAL \citep{OPAL} tables  as representative of uncertainties in opacity
calculations, i.e. we assume
\begin{equation}
  \delta        \kappa_{\rm        opa}(r)       \equiv        \frac{\kappa_{\rm
    OPAL}(\overline{T}(r),\overline{\rho}(r),\overline{Y}(r),\overline{Z}_i(r))}
                {\kappa_{\rm OP}(\overline{T}(r),\overline{\rho}(r),
                  \overline{Y}(r),\overline{Z}_i(r))} - 1 
                \label{dkappaOPALOP}
\end{equation}
The function $\delta \kappa_{\rm opa}(r)$ is shown in Figure~\ref{Fig3}.
The sound speed errors obtained with this assumption are much larger that what
is obtained by the standard  approach, even if $|\delta \kappa_{\rm opa}(r)| \le
0.025$ almost everywhere.

%=============================
\begin{table}[t]
\begin{center}{
\small
\begin{tabular}{l|ccccccccccc} 
\hline \hline &  {\texttt Age} & {\texttt Diffu} & {\texttt  Lum} & $S_{11}$ &
$S_{33}$ & 
  $S_{34}$ & $S_{17}$ & $S_{e7}$ & $S_{1,14}$ & {\texttt Opa} \\ 
\hline
  $Y_{\rm b}$ & -0.001 & -0.012 & 0.002 & 0.001 & 0 & 0.001& 0& 0& 0. & 0.004
  \\ 
$R_{\rm b}$ & -0.0004& -0.0029& -0.0001& -0.0006 & 0.0001& -0.0002& 0& 0&
  0& 0.0014 \\ 
\hline $\Phi_{\rm pp}$ & 0  & -0.002& 0.003 & 0.001  & 0.002 &
  -0.003 & 0 & 0 & 0 & -0.001 \\ 
$\Phi_{\rm Be}$ & 0.003 & 0.022 & 0.014 &  -0.010 & -0.023 & 0.047 & 0 & 0 & 0
  & 0.009 \\ 
$\Phi_{\rm B}$ &0.006& 0.044 &  0.029& -0.025& -0.022& 0.046& 0.075& -0.02& 0&
  0.020 \\ 
$\Phi_{\rm N}$ &0.004& 0.054& 0.018& -0.019& 0.001& -0.003& 0& 0& 0.051& 0.013
  \\ 
$\Phi_{\rm O}$ &0.006& 0.062&  0.024& -0.027& 0.001& -0.002& 0& 0& 0.072&
  0.018 \\ \hline
\end{tabular}
}\end{center}
\vspace{-0.5cm} 
\caption{{\protect The contributions $C_{Q,I}$  to uncertainties in
    theoretical  predictions for helioseismic  observables and  solar neutrino
    fluxes. \label{Tab4} } }
\vspace{0.5cm}
\end{table}
%=============================
%=======================================

Finally, the  total theoretical error for  each observable Q  is calculated by
combining in quadrature all the error contributions, i.e.
\begin{equation}
\sigma^2_{Q,\rm theo} = \sum_{I} C^2_{Q,I} 
\end{equation}
where the sum  extends over the 10 parameters listed in  the beginning of this
section.

%%=============================================================================
%\begin{table}[t]
%\begin{center}{
%%\tiny
%\begin{tabular}{l|cccccccccc}
%%
% & {\texttt Age} &  {\texttt Diffu} & {\texttt Lum} &  $S_{11}$ & $S_{33}$ & $S_{34}$ & $S_{17}$ & $S_{e7}$ & $S_{1,14}$ & $S_{\rm hep}$ \\
%\hline
%$\Delta I/I$  &0.0044 & 0.15 & 0.004 & 0.009 & 0.052 & 0.054 & 0.075 & 0.02 & 0.072 & 0.3
%%
%\end{tabular}
%}\end{center}
%%\vspace{0.1cm} 
%\caption{\em {\protect\small
%\label{Tab3}
%The fractional uncertainties for the input parameters in SSM.}}\vspace{0.4cm}
%\end{table}
%
%% ===================================

\subsection{Observational constraints}
\label{ObsConstraints}

In  the third  column of  Table~\ref{Tab2}, we  give the  observational values
$Q_{\rm  obs}$ for  helioseismic  and solar  neutrino  observables. The  solar
neutrino  fluxes are  obtained  by performing  a  fit to  all available  solar
neutrino data  \citep{BorexinoPC}. Among the various components,  the $^8 {\rm
  B}$  and $^7{\rm  Be}$  neutrino  fluxes are  essentially  determined by  SK
\citep{SK} and SNO \citep{SNO}  and by Borexino \citep{Borexino} respectively,
i.e. by  two independent sets of  experimental data.  We thus  include the two
values
\begin{eqnarray}
\nonumber \Phi_{\rm Be, obs} &=& 4.82\, (1^{+0.05}_{-0.04}) \times 10^{9} {\rm
  cm}^{-2}\, {\rm s}^{-1}  \\ \Phi_{\rm B, obs} &=&  5.00\, (1\pm 0.03) \times
10^{6} {\rm cm}^{-2}\, {\rm s}^{-1}.
\end{eqnarray}
and assume their errors are  uncorrelated.  Generically denoting with $U_Q$ an
uncorrelated fractional error in the observable $Q$, we adopt
\begin{eqnarray}
\nonumber
U_{\rm Be} &=& 0.045\\
U_{\rm B} &=& 0.03 
\end{eqnarray}
for  $\Phi_{\rm  Be}$ and  $\Phi_{\rm  B}$  respectively.   We note  that  the
observational  errors  are  smaller  than  the  uncertainties  in  theoretical
predictions.

The surface helium  abundance and the inner radius  of the convective envelope
are obtained by inversion of helioseismic frequencies. We adopt the values
\begin{eqnarray}
\nonumber
Y_{\rm b, obs} &=& 0.2485 \pm 0.0035 \\
R_{\rm b, obs} &=& 0.713 \pm 0.001
\end{eqnarray}
which  are obtained in  \cite{BasuYs} and  \cite{BasuRb} respectively,  and we
indicate with
\begin{eqnarray}
\nonumber
U_{\rm Y} &=& 0.015 \\
U_{\rm R} &=&  0.0014
\end{eqnarray}
the  fractional   errors  that  we  assume,   as  it  is   usually  done  (see
e.g. \citealp{pinso}), not being significantly correlated.

The  sound speed  data  points $\delta  c_i$  reported in  Figure~1 have  been
obtained  in \cite{basu} with  the set  of frequencies  of solar  low-degree p
modes  from the  BiSON network  by using  the Subtractive  Optimally Localized
Averages (SOLA) inversion  technique. The various points are  localized at the
target  radii $r_i$  of the  corresponding averaging  kernels.   The displayed
error bars  $U_{i, \rm exp}$  are calculated by propagating  the observational
uncertainties of  the measured  frequencies. It is  well known,  however, that
larger  errors  arise from  the  choice of  the  parameters  in the  inversion
procedure and from the assumed starting model for the inversion.  An extensive
investigation of  uncertainties in helioseismic determinations  of sound speed
has  been performed  in  \citet{dziembo}\footnote{\citet{dziembo} adopted  two
  different  inversion   methods,  depending  on  the  value   of  the  radial
  coordinate.  For  $r/R_{\odot}\ge 0.1$,  the Regularized Least  Square (RLS)
  method was  used. In the inner  region, an hybrid method  combining the SOLA
  and the RLS techniques was used.}.
The red band in Figure~\ref{Fig1} corresponds to the so-called ``statistical''
uncertainty $U_{\rm stat}(r)$ that  is obtained in \cite{dziembo} by combining
in quadrature all the relevant error contributions.
In our  analysis, we  define total (fractional)  observational errors  for the
sound speed by combining in quadrature experimental and ``statistical'' errors
according to:
\begin{equation}
U_{c, i} = \sqrt{U_{i,\rm exp}^2 + U^2_{\rm stat}(r_i) }
\end{equation} 
where the index $i$ indicates the considered data point.  Clearly, sound speed
determinations at  different radii  are expected to  have a certain  degree of
correlation  because  uncertainties  are   mainly  related  to  the  inversion
procedure.   Unfortunately,   the  information  provided   in  the  scientific
literature does not allow us  to quantify these correlations. For this reason,
we  include  the  sound  speed  errors as  being  uncorrelated.   The  correct
quantification of helioseismic errors is  an important ingredient and it would
be  desirable   that  the  complete   information  were  provided   in  future
investigations.

As it  is well known,  solar models implementing  the AGSS09 admixture  do not
correctly reproduce  the helioseismic constraints. By  combining in quadrature
theoretical and observational errors, we  see that the predictions for $Y_{\rm
  b}$ and $R_{\rm b}$ deviates from observational data at $\sim 3.6 \, \sigma$
and  $\sim 3.9  \, \sigma$  respectively\footnote{These discrepancies are sligthly
different from what found by \citet{Bahcall06}. This is due to the fact that 
we use a different prescription for the opacity uncertainty;
we do not include composition uncertainties in the error budget;
\citet{Bahcall06} use the surface composition of \citet{AGS05}.}. The  sound speed  at $r\sim  0.65 \,
R_\odot$  differs from  the results  of  helioseismic inversion  by $\sim  4.6
\,\sigma$.   Helioseismic  data  are   much  better  fitted  by  solar  models
implementing  the   GS98  surface  composition,   as  it  can  be   seen  from
Table~\ref{Tab2}  and Figure~\ref{Fig1}.  A  reasonable agreement  exists between
predicted and reconstructed solar neutrino  fluxes both for the AGSS09 and the
GS98 solar models.

%%%%%%%%%%%%%%%%%%%%%%%%%%%%%%%%%%%%%%%%%%%%%%%%%%%%%%
%%%%
%%%% Statistical approach
%%%% 
%%%%%%%%%%%%%%%%%%%%%%%%%%%%%%%%%%%%%%%%%%%%%%%%%%%%%%
\section{The statistical approach}
\label{sec:stat}

Our goal is to build a $\chi^2$ function that can be used as a figure-of-merit
for SSMs  with different  compositions. Let us  consider a  generic observable
quantity with its associated observational value $Q_{\rm obs}$ and theoretical
prediction $Q$.  We indicate with
\begin{equation}
\delta Q_{\rm obs} = \frac{Q_{\rm obs}}{Q}-1
\end{equation}
the fractional difference between  the observational value and the theoretical
result.  In  this work,  we consider  a set of  $N=34$ differences  $\delta Q$
given by
\begin{equation}
\{\delta Q_{\rm obs} \} = \{ \delta\Phi_{\rm B},\, \delta \Phi_{\rm Be},\, \delta Y_{\rm
  b},\,  \delta R_{\rm  b};\, \delta  c_{1},\, \delta  c_{2},  \dots,\, \delta
c_{30} \} 
\end{equation}
where we  include the sound  speed determinations $c_{i,\rm obs}$ of  \cite{basu} that
are localised at $r\le 0.8 R_\odot$.

The differences $\delta Q_{\rm obs}$ are affected by the uncorrelated errors 
$U_Q$ (e.g.  from neutrino experiments and  helioseismic data) and by a set of
systematic correlated errors $C_{Q,I}$  induced by $K=10$ independent sources
that, in our approach, are\footnote{In this work, correlated systematic errors incidentally coincide 
with theoretical uncertainties, while uncorrelated errors coincide with observational uncertainties. 
However, this is not necessarily the case. For  example,  were  we to  use  $\Phi_{\rm  pp}$ as  an
  observable  in our  analysis, there  would exist  a correlation  between its
  experimental error and  that of $\Phi_{\rm Be}$ due  to the \emph{luminosity
    constraint} usually  adopted in the  analysis of solar neutrino  data, see
  e.g. \citet{maltoni:10}.}
\begin{equation}
\{ I \} = \{ \texttt{opa};\, \texttt{age}; \,\texttt{diffu};\, \texttt{lum};\,
S_{11},\, S_{33},\, S_{34},\, S_{17},\, S_{e7},\, S_{1,14} \} 
\end{equation}
Following \cite{lisi}, we define $\chi^2$ as
%...........................................................................
\begin{equation}\label{chi2}
\nonumber
\chi^2 = \min_{\{\xi_I\}}\left[\sum_{Q}
\left(\frac{\delta Q_{\rm obs} -
\sum_{I}\xi_I \, C_{Q,I}}{U_Q}\right)^2+\sum_{I}
\xi_I^2\right].
\end{equation}
%...........................................................................
This definition describes the effects of systematic correlated errors $C_{Q,I}$ by introducing the shifts $-\xi_I \, C_{Q,I}$, where $\xi_I$
is a univariate gaussian random variable. Expressing $\chi^2$ in this way 
is completely equivalent to the standard covariance matrix approach (for
a formal 
proof refer to \citealt{lisi}).  However, it offers some relevant advantages:
1) it  is more easily implemented numerically  and; 2) it allows  to trace the
individual contributions  to the $\chi^2$.  Denoting  with $\tilde{\xi}_I$ the
values that minimize the $\chi^2$, one obtains
\begin{equation}
\chi^2 \equiv  \chi^2_{\rm obs}+\chi^2_{\rm  syst} = \sum_{Q}  \tilde{X}_Q^2 +
\sum_{I} \tilde{\xi}_I^2 
\end{equation}
where
\begin{equation}
\tilde{X}_Q\equiv   \frac{\delta  Q_{\rm   obs}  -   \sum_{I}\tilde{\xi}_I  \,
  C_{Q,I}}{U_Q} 
\label{PullObs}
\end{equation}
are  the   so-called  ``pulls''   of  observational  quantities.   The  values
$\tilde{\xi}_I$ are instead referred to as the ``pulls'' of systematical error sources
that, in our analysis, coincides with input parameters in solar model construction.
The  distribution of the $\tilde{\xi}_I$  can thus be used to highlight tensions in SSM assumptions.
The optimal  composition of Sun  is found by  minimizing the $\chi^2$  and the
obtained value $\chi^2_{\rm min}$ provides  information on the goodness of the
fit. The allowed regions are determined by cutting at prescribed values of the
variable $\Delta  \chi^2 \equiv \chi^2  - \chi^2_{\rm min}$.

%=============================================================================
\begin{table}[t]
\begin{center}{
\small
\begin{tabular}{l|cccccccc|cc} 
\hline
\hline
  & C & N & O & Ne & Mg & Si & S & Fe & CNO & Met \\ \hline
$Y_{\rm b}$ &  -0.003 & 0.001 & 0.025 & 0.030 & 0.032 & 0.063 & 0.043 & 0.086 & 0.023& 0.223 \\
$R_{\rm b}$ & -0.005 & -0.003 & -0.027 &  -0.011 &  -0.004 &    0.002 & 0.004 &  -0.009 & -0.035 & -0.007  \\
\hline
$\Phi_{\rm pp}$ & -0.005 & -0.001 &  -0.004 & -0.004 & -0.004 &  -0.008  &  -0.006 &  -0.017  &  -0.010 & -0.034 \\ 
$\Phi_{\rm Be}$ &  0.004 &  0.002  & 0.052 & 0.046  & 0.048 & 0.103 & 0.073 & 0.204 & 0.058 & 0.429 \\
$\Phi_{\rm B}$ & 0.026 & 0.007 & 0.112 & 0.088 & 0.089 & 0.191 & 0.134 &  0.501 & 0.145 & 0.916 \\
$\Phi_{\rm N}$ & 0.874 & 0.147 &  0.057 & 0.042 & 0.044 & 0.102  & 0.072 & 0.263 & 1.078 & 0.480 \\ 
$\Phi_{\rm O}$ &  0.827 & 0.206 & 0.084 & 0.062 & 0.065 & 0.145 & 0.102 & 0.382  & 1.117 & 0.694 \\ \hline
\end{tabular}
}\end{center}
\vspace{-0.5cm} 
\caption{\protect The logarithmic derivatives ${\mathcal B}_{Q,j} =
    d   \ln   Q   /   d   \ln   z_{j}  $   with   respect   to   the   surface
    abundances. \label{LogDevComp}}
\vspace{0.5cm}
\end{table}

% ====================================================

\section{Describing the role of metals}
\label{sec:metals}

We study  the response of  the Sun to  changes in the heavy  element admixture
$\left\{z_{\rm j}\right\}$, expressed in terms of the quantities
\begin{equation}
z_{\rm j} \equiv Z_{\rm j,b} /X_{\rm b}
\end{equation}
where $Z_{\rm j, b}$ is the surface  abundance of the $j-$element, $X_{\rm b}$ is that of
hydrogen,   and  the   index  $j$   runs   over  all   relevant  metals   (see
Table~\ref{LogDevComp}).  We  determine the  dependence of the  observables on
the surface composition by constructing solar models in which each $z_{\rm j}$
is varied individually. We observe that  the effects produced by  a change of
composition $\left\{\delta z_{j}\right\}$ are well described
by a linear relation
\begin{equation}
\delta Q = \sum_j{{\mathcal B_{Q,\rm j}} \; \delta z_{\rm j}},
\label{linear}
\end{equation}
%between the fractional variation of the $Q$ observable: 
%\begin{equation}
%\delta Q \equiv \frac{Q}{\overline{Q}} -1
%\end{equation}
where $\delta z_{\rm j}$ is the fractional variations of $z_{\rm j}$
\begin{equation}
\delta z_{\rm j} \equiv \frac{z_{\rm j}}{\overline{z}_{\rm j}} - 1
\end{equation}
with  respect  to the  AGSS09  value  $\overline{z}_{\rm  j}$.  In  this
assumption,   the  coefficients   ${\mathcal  B}_{Q,\rm   i}$   represent  the
logarithmic  derivatives  of  $Q$  with  respect to  the  $j-$element  surface
abundance, i.e.
\begin{equation}
{\mathcal B}_{Q,\rm j} = \frac{d \ln Q}{d \ln z_{\rm j}}.
\end{equation}
%

\begin{figure}[t]
\par
\begin{center}
\includegraphics[width=8cm,angle=0]{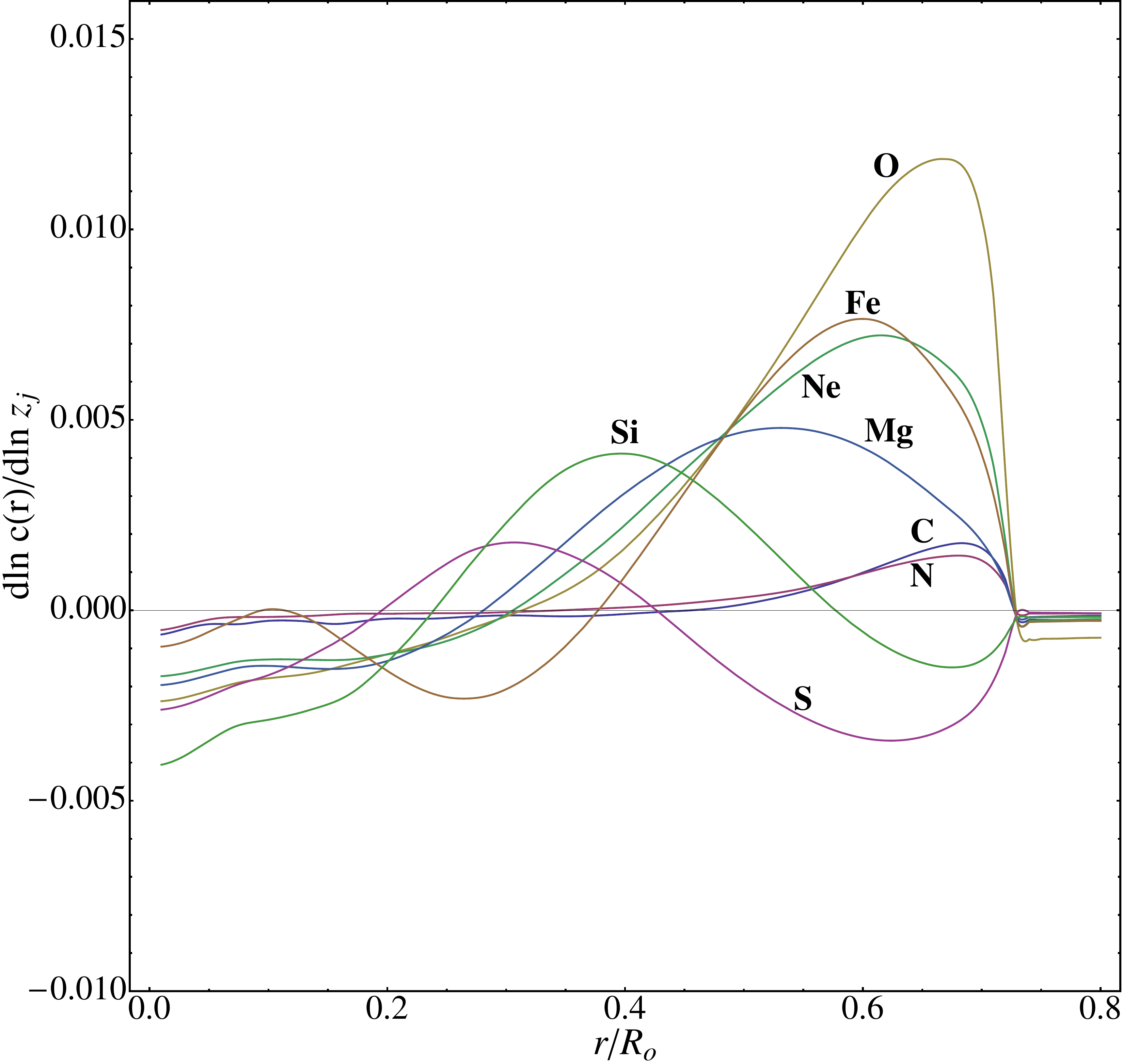}
\includegraphics[width=8cm,angle=0]{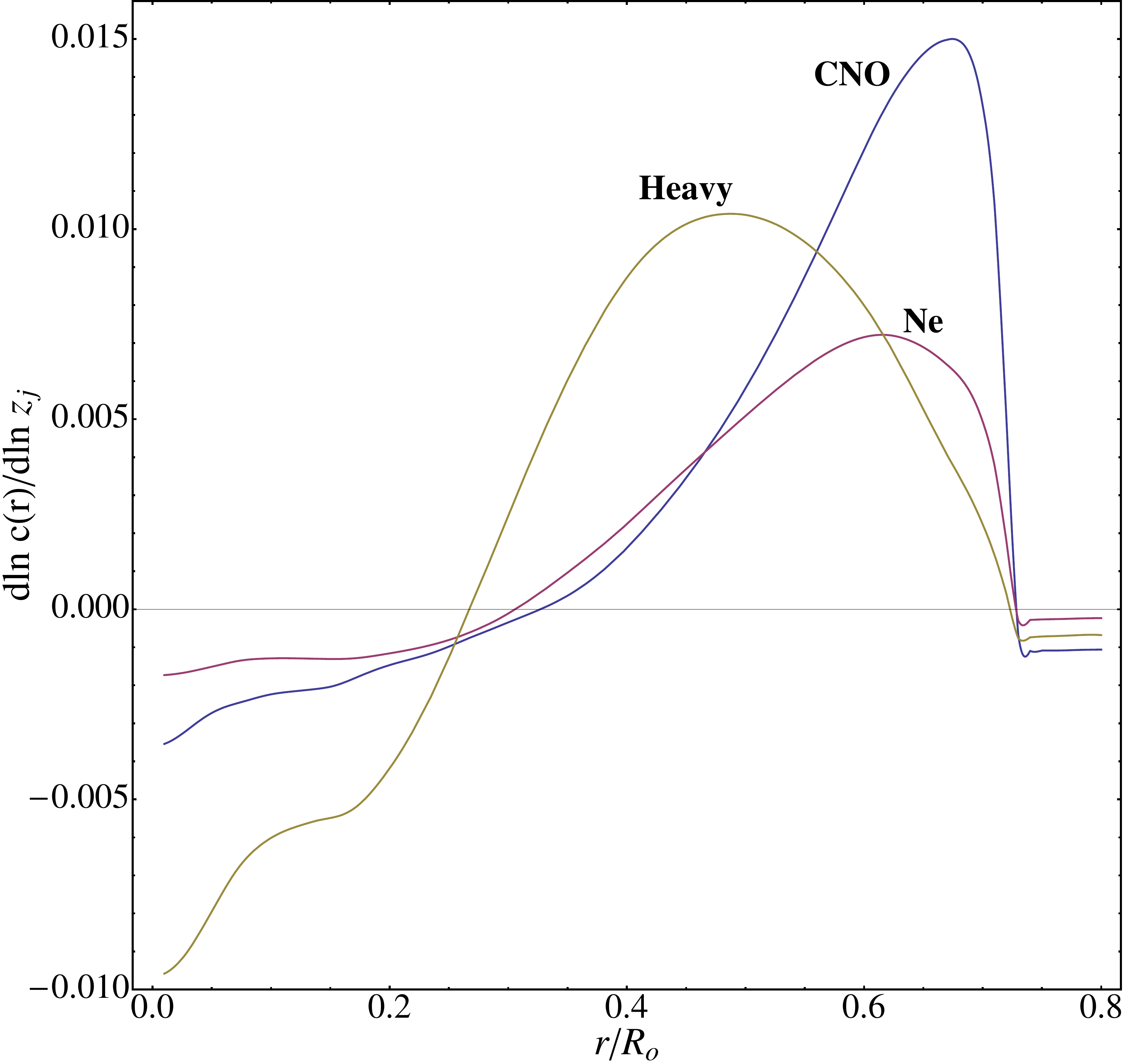}
\end{center}
\par
\vspace{-5mm}   \caption{\protect  Left   panel:  Logarithmic
  derivatives of the sound speed with respect to the surface abundances $z_j$.
  Right Panel:  Logarithmic derivatives of the sound  speed with respect
  to  the total  ${\rm CNO}$,  ${\rm  Ne}$ and  \emph{meteoritic} elements  surface
  abundances.}
\label{LogDevCsComp}
\end{figure}

%\textbf{This needs to be checked}
The values obtained for the ${\mathcal B}_{Q,\rm j}$ coefficients are reported
in Table~\ref{LogDevComp}.   Our results can be compared  with the coefficients
presented in  \cite{serenelli:2009}. The small differences arise  from the fact
that we considered relatively large  variations for the various abundances, in
order  to  check  the  adequacy  of rel.~(\ref{linear})  over  the  ranges  of
compositions required by our  analysis.  The logarithmic derivative ${\mathcal
  B}_{\rm c, j}(r)$ of the sound speed with respect to the surface composition
have not  been shown elsewhere in  scientific literature and are  given in the
left panel of Figure~\ref{LogDevCsComp}.

\begin{figure}[t]
\par
\begin{center}
\includegraphics[width=10cm,angle=0]{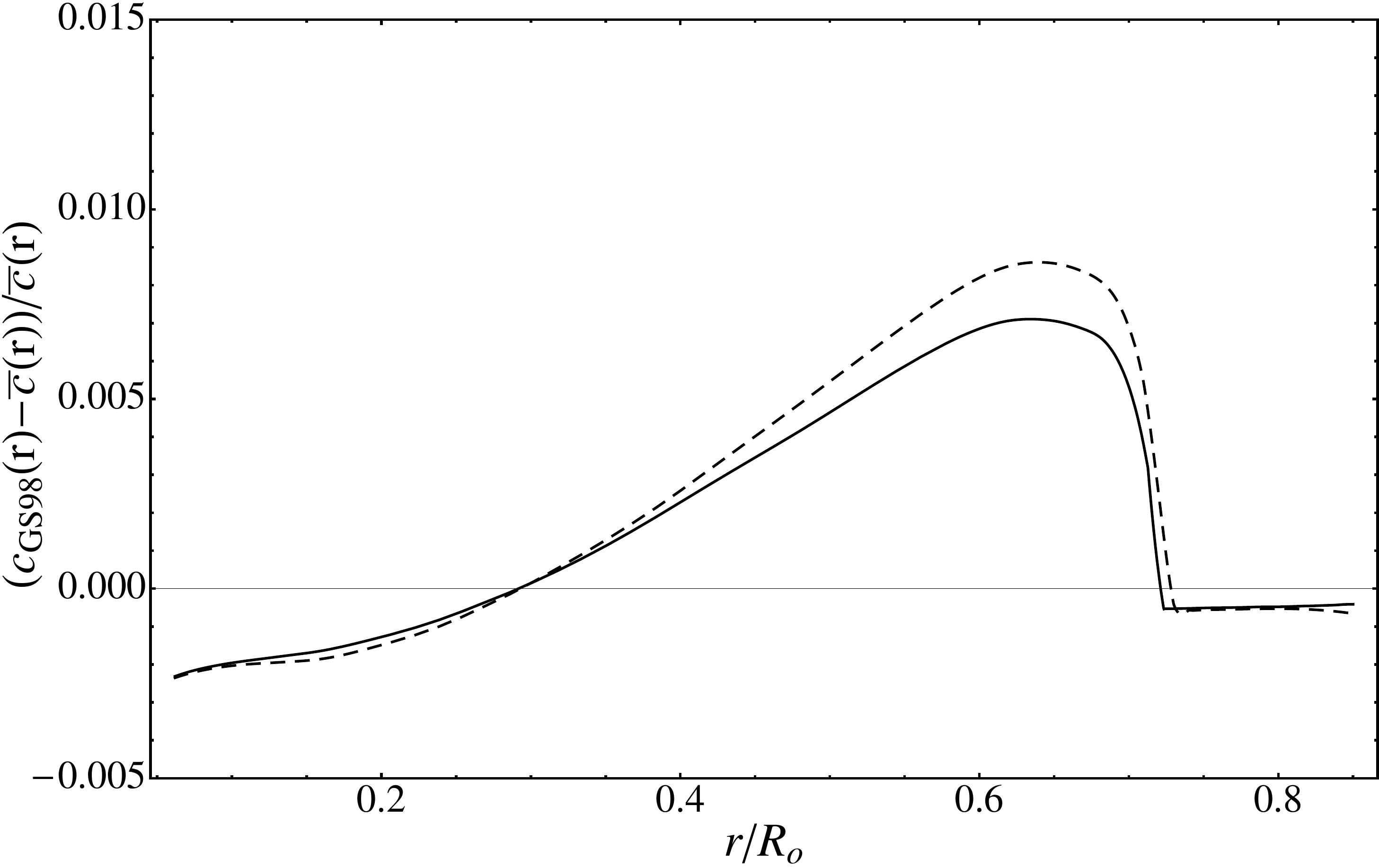}
\end{center}
\par
\vspace{-5mm} \caption{\protect The solid line  shows the fractional
  difference  between the  sound  speed predicted  by  SSMs implementing  GS98
  admixture and that obtained by using AGSS09 surface compositions. The dashed
  line  is  obtained   by  using  the  linear  expansion   given  by  equation
  (\ref{linear}).} 
\label{LinearTest}
\end{figure}

The accuracy of equation~(\ref{linear}) can be tested by using it to reproduce
the  predictions of  the  GS98 solar  model  starting from  those obtained  by
implementing   the  AGSS09   surface   composition.   The   results  of   this
exercise, the \emph{reconstructed} GS98 (GS98rec) model, are
shown in  the last column of Table~\ref{Tab2}  and in Figure~\ref{LinearTest}.
The  differences between  GS98 and  AGSS09 are  reproduced with  $\sim  20 \%$
accuracy (in  the worst  cases) and the  errors introduced  by the use  of the
linear  approximation are  always  smaller or  comparable  to theoretical  and
observational  uncertainties.   The  use  of  relation~(\ref{linear})  greatly
simplifies  the numerical  problem  of  scanning over  the  range of  possible
compositions.   In this  assumption, indeed,  the $\chi^2$  is expressed  as a
quadratic  function of  the various  $\delta  z_{j}$ that  can be  effectively
minimized.  We obtain
%...........................................................................
\begin{equation}\label{chi2_2}
\nonumber
\chi^2 = \min_{\{\xi_I\}}\left[\sum_{Q}
\left(\frac{\delta \overline{Q} - 
\sum_{j} \delta z_{j} \, {\mathcal B}_{Q, j} -
\sum_{I}\xi_I \, C_{Q,I}}{U_Q}\right)^2+\sum_{I}
\xi_I^2\right]\ .
\end{equation}
%...........................................................................
where
\begin{equation}
\delta \overline{Q} = \frac{Q_{\rm obs}}{\overline{Q}} - 1
\end{equation}
is the fractional difference between the observational value $Q_{\rm obs}$ and
the value $\overline{Q}$ predicted by  the AGSS09 solar model.  Even with this
simplification, however, it  is not possible (nor useful)  to consider all the
$\delta  z_{j}$  as free  parameters.  For  this reason,  we  group metals
according to the  method by which their abundances  are determined.  Following
\cite{delahaye:2010}, we consider three different groups given by $({\rm C + N
  + O})$, ${\rm Ne}$,  $({\rm Mg + Si +S + Fe})$  which include elements whose
abundances are determined in the  photosphere, in the chromosphere and corona,
and in the meteorites, respectively.  Within each group, we vary the elemental
abundances according to same multiplicative factor.  In other words, we define
three independent  parameters ($\delta z_{\rm  CNO},\delta z_{\rm Ne},\;\delta
z_{\rm met}$) as
\begin{eqnarray}
\nonumber
1+\delta  z_{\rm CNO} &\equiv&  \frac{z_{\rm C}}{\overline{z}_{\rm  C}} \equiv
\frac{z_{\rm      N}}{\overline{z}_{\rm      N}}      \equiv      \frac{z_{\rm
    O}}{\overline{z}_{\rm O}}\\ 
\nonumber
1+\delta z_{\rm Ne} &\equiv& \frac{z_{\rm Ne}}{\overline{z}_{\rm Ne}}  \\
\nonumber
1+\delta  z_{\rm  met}  &\equiv&  \frac{z_{\rm  Mg}}{\overline{z}_{\rm  Mg}}
\equiv   \frac{z_{\rm    Si}}{\overline{z}_{\rm   Si}}   \equiv   \frac{z_{\rm
    S}}{\overline{z}_{\rm S}} \equiv \frac{z_{\rm Fe}}{\overline{z}_{\rm Fe}} 
\end{eqnarray}
Based  on   this  assumption  and  the   linear  relation~(\ref{linear}),  the
logarithmic derivatives ${\mathcal B}_{Q,\rm CNO}(r)$ and ${\mathcal 
  B}_{Q,\rm met}(r)$ of the various  observables $Q$ with respect to $\delta
z_{\rm CNO}$ and $\delta z_{\rm met}$ are
\begin{eqnarray}
\nonumber
{\mathcal B}_{Q,\rm CNO} &\equiv& {\mathcal B}_{Q,\rm C} + {\mathcal B}_{Q,\rm
  N} + {\mathcal B}_{Q,\rm O}\\ 
\nonumber
{\mathcal  B}_{Q,\rm  met}  &\equiv&  {\mathcal B}_{Q,\rm  Mg}  +  {\mathcal
  B}_{Q,\rm Si}+ {\mathcal B}_{Q,\rm S} + {\mathcal B}_{Q,\rm Fe} 
\end{eqnarray}
and are reported in Table~\ref{LogDevComp}.  The functions ${\mathcal B}_{c,\rm
  CNO}(r)$, ${\mathcal  B}_{c,\rm Ne}(r)$ and  ${\mathcal B}_{c,\rm met}(r)$
  describing the effects of each group of elements on the sound speed
profile of the Sun are shown in the right panel of Figure~\ref{LogDevCsComp}.

\section{Inferring the solar composition}
\label{sec:results}

\subsection{Volatiles and refractories: a two-parameter analysis}

As a  first application,  we consider a  scenario in which  the neon-to-oxygen
ratio is  fixed to  the value  prescribed by the  AGSS09 compilation,  i.e. we
further constrain  the possible variations  of the heavy element  admixture by
assuming $\delta  z_{\rm CNO}=  \delta z_{\rm Ne}$.   In this  hypothesis, the
$\chi^2$  is defined  in  terms of  two  independent parameters  $\left(\delta
z_{\rm  CNO}  \,  ,\delta  z_{\rm   met}\right)$  that  are  varied  to  fit
helioseismic  and solar neutrino  constraints.  A similar exercise was performed 
in \cite{pinso} where, however, only the determinations of the surface helium 
abundance and of the convective radius were considered. 
Here, we include the information provided by the sound speed profile and 
the neutrino fluxes in a global quantitative analysis.
The redundancy of the different pieces of experimental information allows us
to obtain more solid constraints on the solar compositon and, even more relevant, 
to verify that a coherent picture emerge from the data and/or
to highligth tensions in SSM assumptions.

\begin{figure}[t]
\par
\begin{center}
\includegraphics[width=8cm,angle=0]{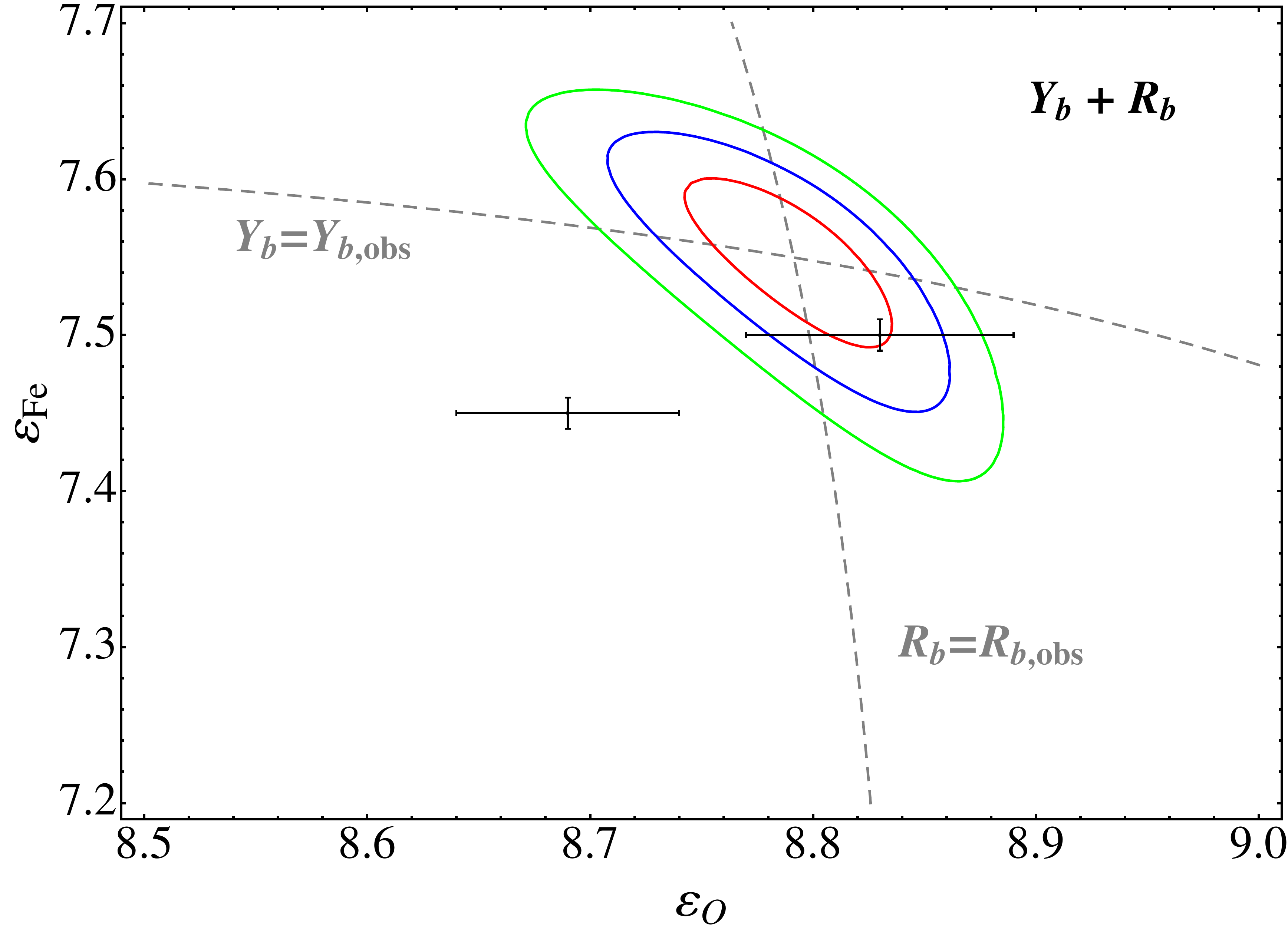}
\includegraphics[width=8cm,angle=0]{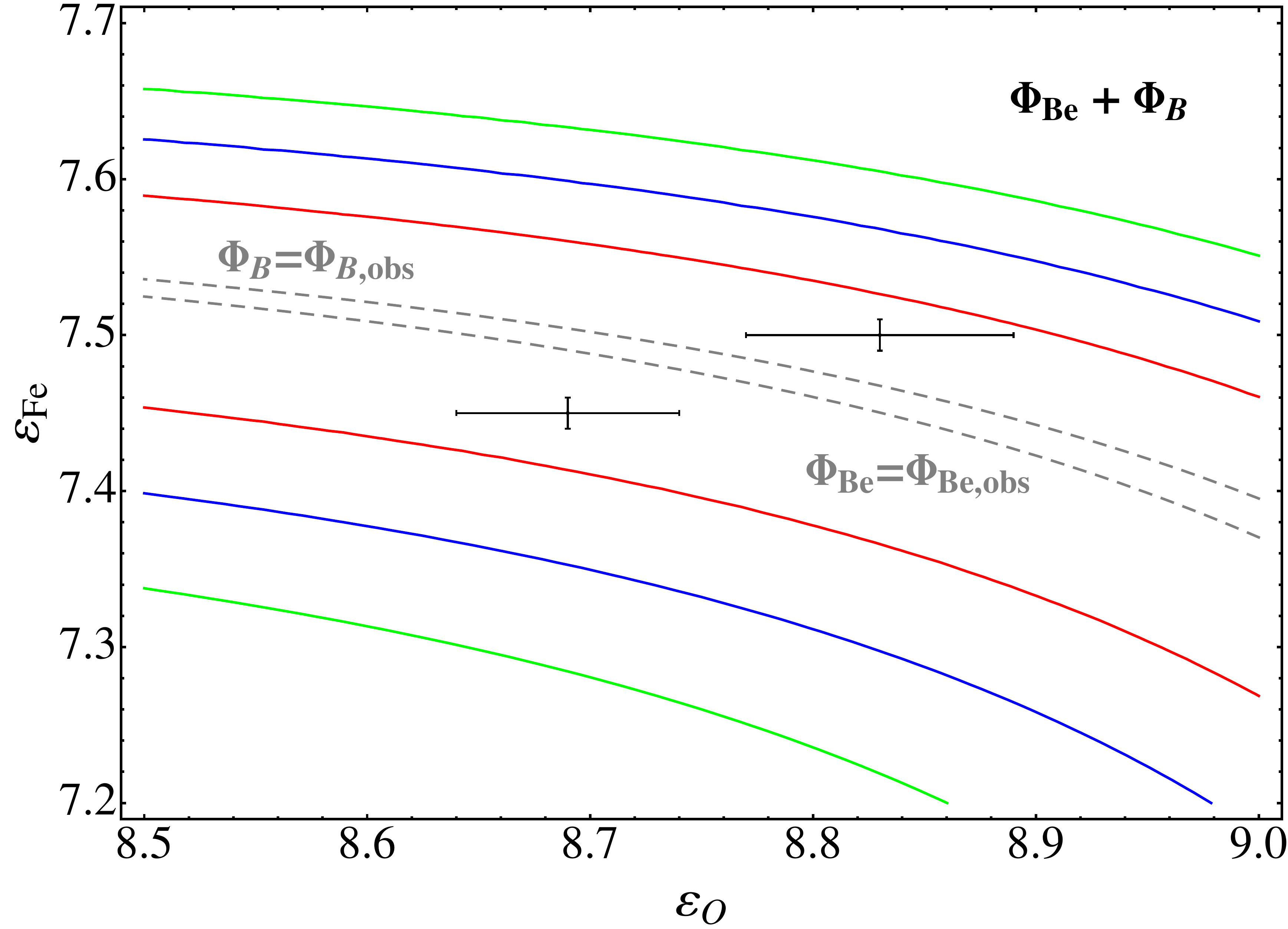}
\includegraphics[width=8cm,angle=0]{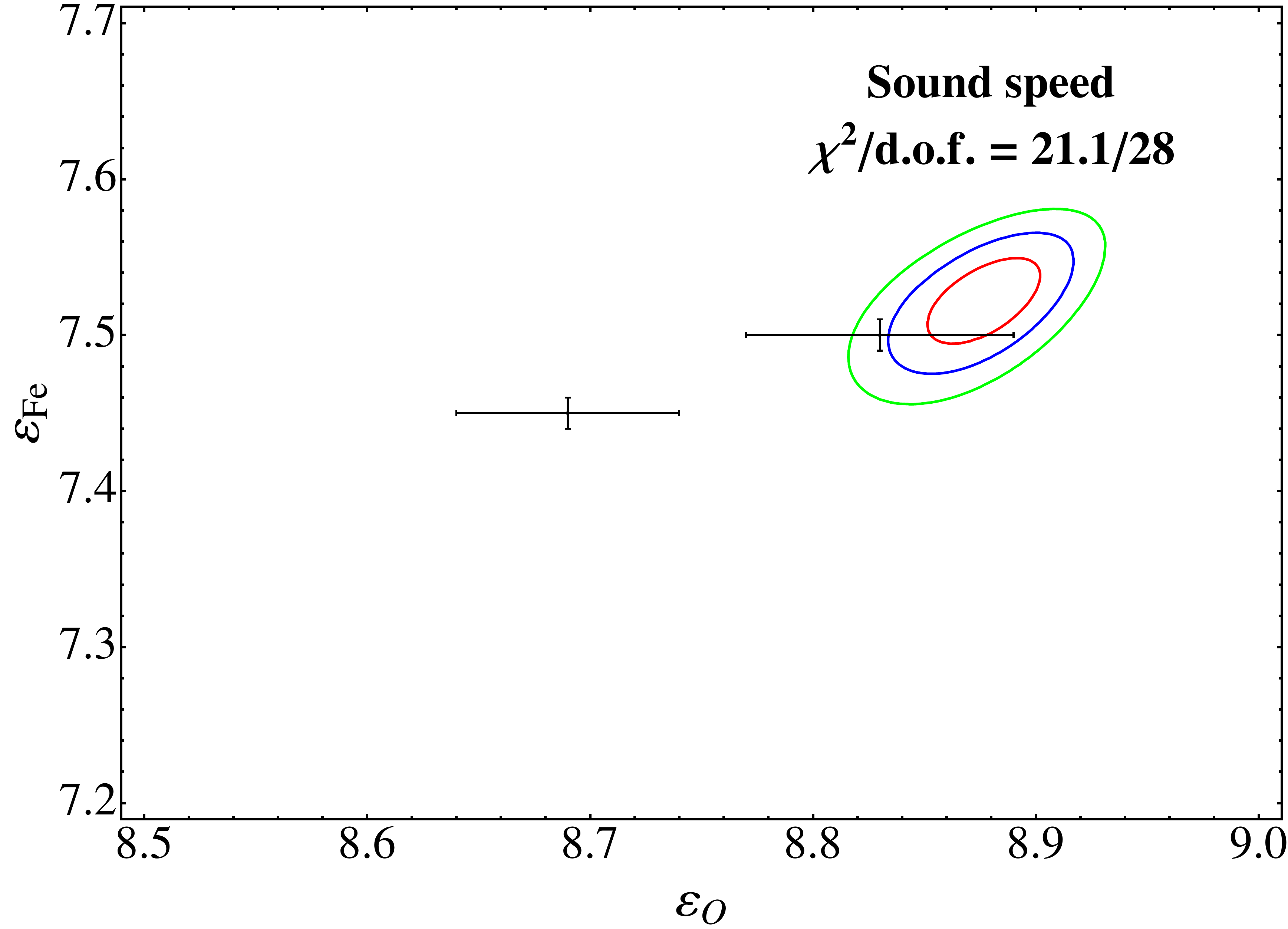}
\includegraphics[width=8cm,angle=0]{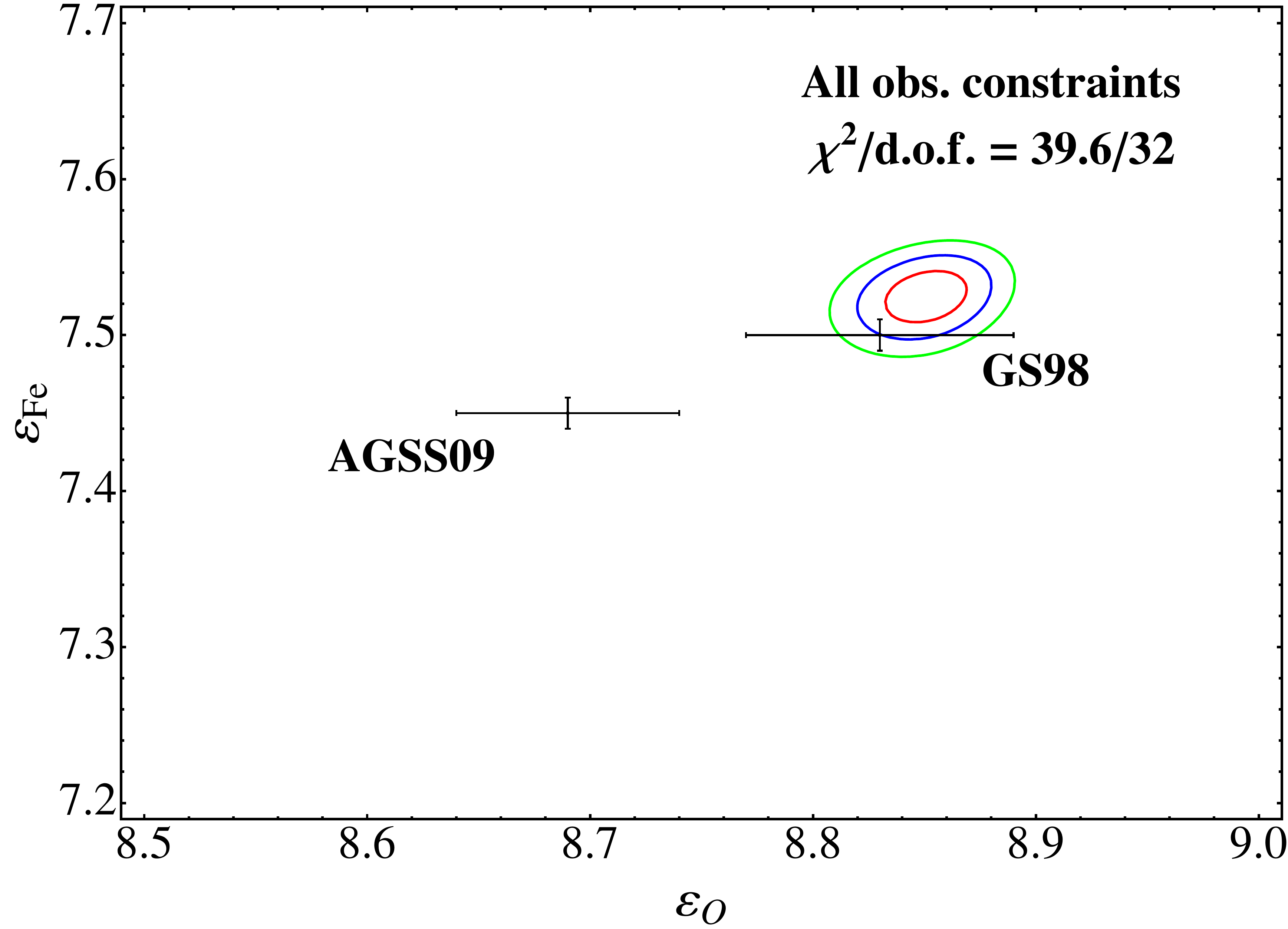}
\end{center}
\par
\vspace{-5mm} \caption{\protect  The bounds on  $\varepsilon_{\rm O}$
  and $\varepsilon_{\rm Fe}$ that are obtained from observational constraints.
  See text for details.}
\vspace{0.5cm}
\label{results2par}
\end{figure}

Our results are  presented in
Figure~\ref{results2par}  where we  use the  astronomical scale  for logarithmic
abundances  $\varepsilon_j$  in  order   to  facilitate  the  comparison  with
observational data.  The conversion from $\delta z_{j}$  to $\varepsilon_j$ is
obtained by using the relation
\begin{equation}
\varepsilon_{j} = \overline{\varepsilon}_{j}+\log{\left(1+\delta z_{j}\right)}
\end{equation}
with   the    AGSS09   abundances   $\overline{\varepsilon}_{j}$    given   in
Table~\ref{Tab1}.
The  coloured   lines  are  obtained   by  cutting  at   $\Delta  \chi^2\equiv
\chi^2-\chi^2_{\rm  min} = 2.3,\,  6.2,\, 11.8$  that correspond  to $1,\,2,\,
3\,\sigma$ confidence levels for a  $\chi^2$ variable with 2 d.o.f..  The data
points show the observational values for the oxygen and iron abundances in the
AGSS09 and GS98 compilations. In  order   to  show  how  different  observational   information  combine  in
determining the optimal composition, we present separately the bounds obtained
by using the: the helioseismic constraints on the surface helium abundance and
the convective radius  (upper-left panel); the $^7{\rm Be}$  and $^{8}{\rm B}$
neutrino  flux determinations  (upper-right panel);  the 30  sound  speed data
points $c_{i,\rm obs}$ from  \cite{basu} that are localized at  $r\le 0.8\, R_{\odot}$
(lower-left  panel); all  the observational  data  simultaneously (lower-right
panel). The main conclusions of our analysis are discussed in the following

\begin{enumerate}

\item [1)] The SSM implementing AGSS09 composition is excluded at a high
confidence being $\chi^2/{\rm  d.o.f.}= 176.7/32$ when all the available
observational constraints are considered.  This result essentially arises from
helioseismic  observables   that  are  in  severe   disagreement  with  AGSS09
predictions.   The  $^7{\rm  Be}$   and  $^8{\rm   B}$  solar   neutrino  flux
determinations  do  not discriminate  among  different  compositions with  the
sufficient level of accuracy. This  is mainly due to theoretical uncertainties
which  are  dominated  by   the  contributions  from  $S_{34}$  and  $S_{17}$,
respectively, but also to a very mild dependence on CNO abundances.

\item  [2)] There is  a reasonable agreement between  the information
  provided  by the various  observational constraints,  as it  can be  seen by
  comparing the different panels of Figure~\ref{results2par}.  The best fit to
  the observational data is obtained for:
\begin{eqnarray}
\nonumber
\delta z_{\rm CNO} = \delta z_{\rm Ne} = 0.45 \pm 0.04 \\
\delta z_{\rm met} = 0.19 \pm 0.03
\label{bestfit2par}
\end{eqnarray}
that correspond  to $\varepsilon_{\rm O}=8.85 \pm  0.01$ and $\varepsilon_{\rm
  Fe}=7.52 \pm  0.01$.  These  values
%, which are close to those obtained 
%by \citet{pinso},  
are consistent  at $\sim  1\sigma$ level
with those  quoted in the GS98 compilation.   
They are close to the results of \citet{pinso}
but have considerably smaller uncertainties.
The quality of the  fit is quite
good  being  the $\chi_{\rm  min}^2/{\rm  d.o.f.}= 39.6  /  32$  when all  the
observational  constraints  are  considered.    The  errors  on  the  inferred
abundances $\varepsilon_{\rm  O}$ and $\varepsilon_{\rm Fe}$  are smaller than
what is obtained  by observational determinations. One caveat  is, however, that we
are considering a simplified  scenario in which different elemental abundances
are grouped together and forced to vary by the same multiplicative factors.
 
\begin{table}[t]
\begin{center}{
\small
\begin{tabular}{l|cccccccccc} \hline \hline
&  {\texttt Opa} & {\texttt Age} &  {\texttt Lum} & {\texttt Diffu} & $S_{11}$ &
 $S_{33}$ & $S_{34}$ & $S_{e7}$ & $S_{17}$ & $S_{1,14}$ \\
\hline
$\tilde{\xi}_I$ & 1.07 & 0.03 & -0.41 & -0.74 & $\sim 0$  & 0.46 & -0.97 & 0.32 & -1.20 & $\sim 0$  \\ 
$\tilde{\xi}_I\delta I $ &  1.07 $\left(\kappa_{\rm OPAL}/\kappa_{\rm OP}-1\right)$& $\sim 0$ & -0.0016 & -0.11 &  $\sim 0$ &  0.024 & -0.05 & 0.007 & -0.09 &  $\sim 0$ \\
\hline
\end{tabular}
}\end{center}
\vspace{-0.5cm} 
\caption{\protect The pulls of systematics $\tilde{\xi}_{I}$ at the
best  fit   point and the fractional variations $\tilde{\xi}_I\delta I$ of the corresponding input parameters.  
All   the  available  observational   information  are  simultaneously fitted. The entries with $\sim 0$ are smaller (in magnitude) than 
$10^{-2}$ in the first line and $10^{-3}$ in the second line.\label{pulls}}
\vspace{0.5cm}
\end{table}

\item   [3)]   The  observational   and   systematic  contribution   to
  $\chi^2_{\rm min}$ are given by  $\chi^2_{\rm obs} = 35.1 $ and $\chi^2_{\rm
  syst} = 4.5$  respectively, with
  the distribution of systematic pulls 
  $\tilde{\xi}_I$ at  the best fit  point reported in  Table~\ref{pulls}.  The
  effects of  systematic pulls (that  correspond to correlated  error sources)
  are relevant and cannot be neglected.   This is seen e.g.  in the upper-left
  panel of Figure~\ref{results2par}  where we see that the  error ellipse axes
  do  not coincide  with  the lines  $Y_{\rm  b}=Y_{\rm b,  obs}$ and  $R_{\rm
    b}=R_{\rm b,  obs}$, as  it would be  expected if error  correlations were
  negligible.  It is also evident from the red dots and the red line in
  Figure~\ref{sound}   that  show   the  predictions   $Q$  of   solar  models
  implementing  the  best  fit  composition  calculated by  using  the  linear
  expansion~(\ref{linear}). 
%  without  taking   into  account   the   pulls  of  systematics. 
These predictions deviate from the observational constraints by
  amounts that are larger than the uncorrelated observational errors. However,
  this does  not imply that the  quality of the fit  is bad since  one has the
  possibility to change the SSM inputs within their range of uncertainty as it
  is described in equation~(\ref{chi2_2}).  The blue dots and the blue line in
  Figure~\ref{sound} show the quantities
\begin{equation}
\tilde{Q} = Q \left[ 1 + \sum_{I} C_{Q,I} \,\tilde{\xi}_I \right]
\label{Qtilde}
\end{equation}
that  include  the effects  due  to the  pulls  of  systematic errors.   The
quantities $\tilde{Q}$  differ quite substantially from  the corresponding $Q$
and  agree  quite  well  with  the  observational  constraints.  The  dominant
contributions to systematic shifts are shown by the black arrows in Fig.~\ref{sound}
and are provided by:  opacity for the sound speed $c(r)$ (see discussion in the next paragraph);
diffusion coefficients which are decreased by $\sim 11 \%$
in order to improve consistency between $Y_{\rm b}$  and $R_{\rm  b}$ and the sound speed profile $c(r)$; 
the astrophysical factors $S_{34}$ and  $S_{17}$ that are decreased by $\sim 5\%$ and $\sim 9\%$ respectively
in order to improve agreement with $\Phi_{\rm B}$ and $\Phi_{\rm Be}$ measurements.

\begin{figure}[t]
\par
\begin{center}
\includegraphics[height=4.6cm,angle=0]{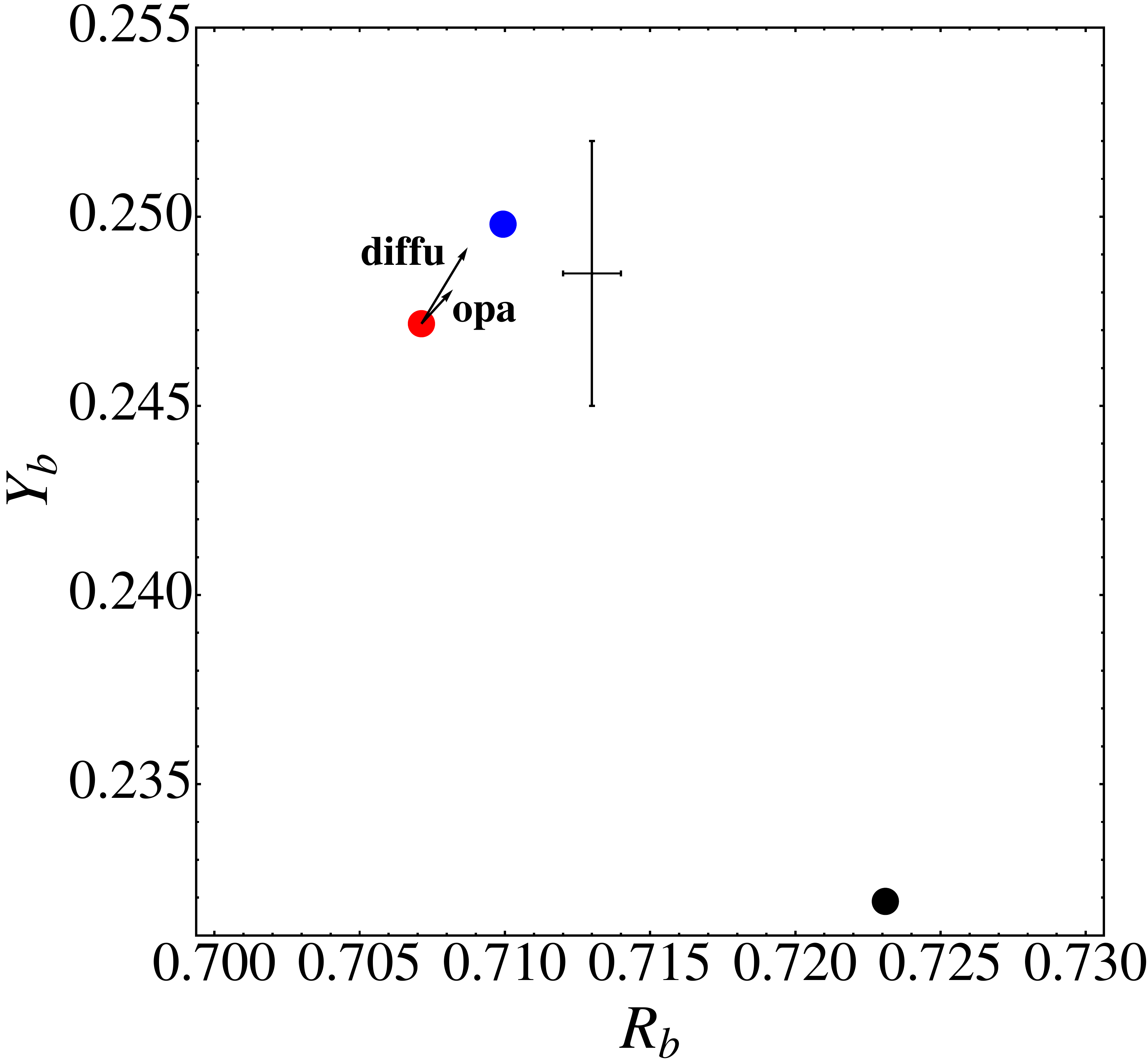}
\includegraphics[height=4.6cm,angle=0]{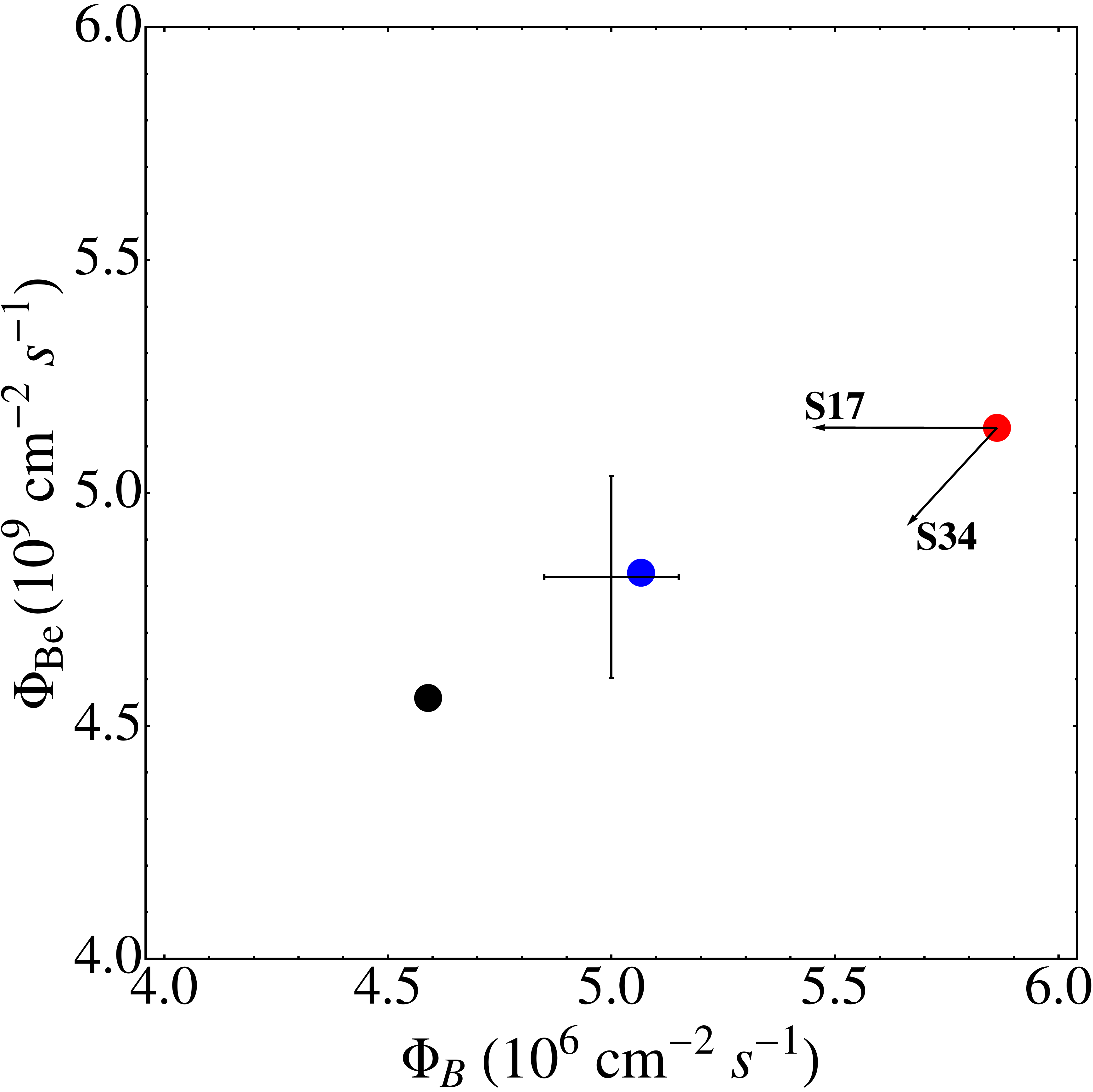}
\includegraphics[height=4.6cm,angle=0]{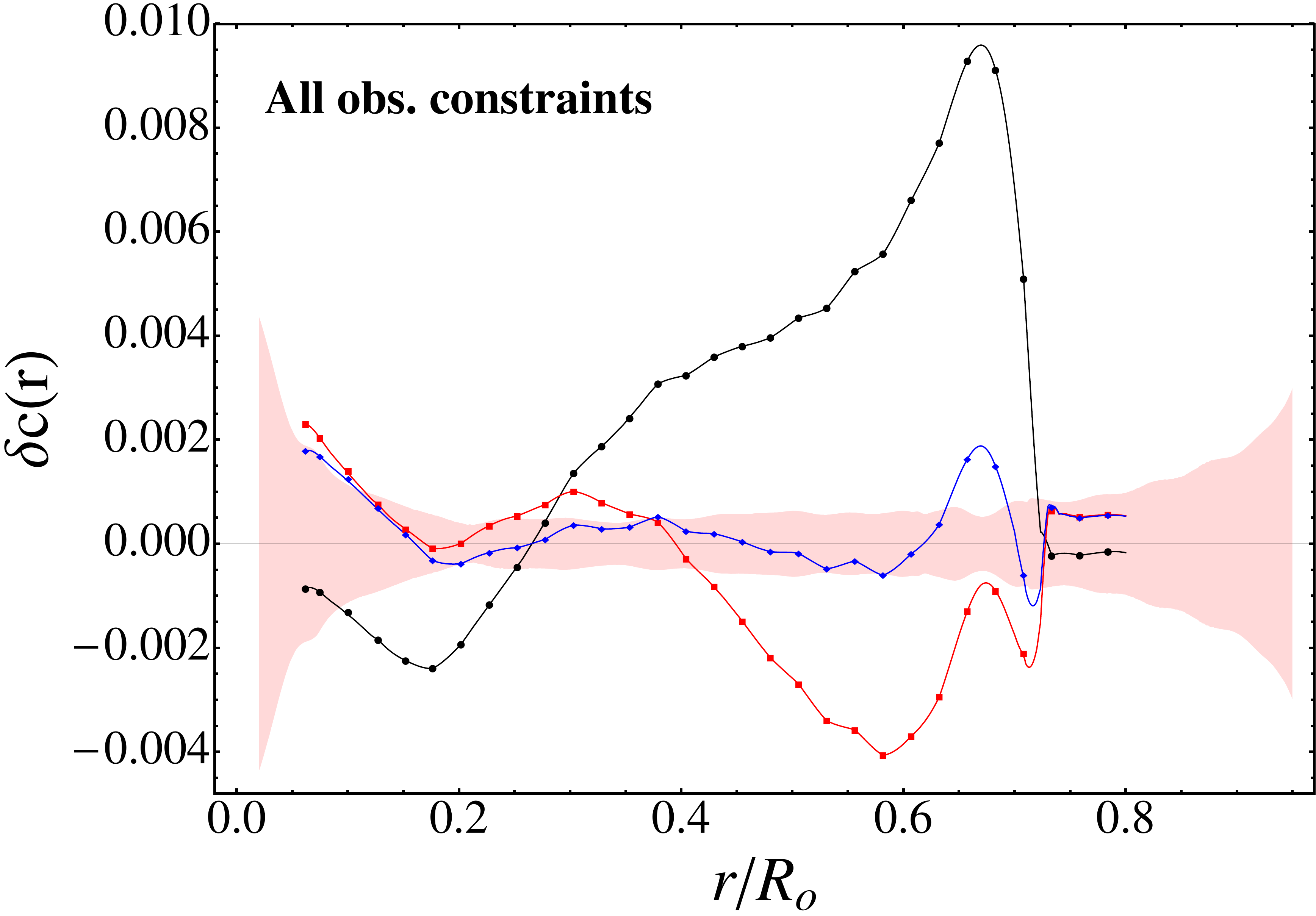}
\end{center}
\par
\vspace{-5mm} \caption{\protect  The helioseismic and  solar neutrino
  observables predicted by  AGSS09 solar model (black) and  by the solar model
  providing  the best  fit to  all observational  constraints with  (blue) and
  without (red) taking into account the pulls of systematical errors.
\label{sound}}
\vspace{0.5cm}
\end{figure}

\item[4)] The large systematic shift of the sound speed due to opacities, $\tilde{\xi}_{\rm opa}=1.07$,
%equation~(\ref{Qtilde}) are dominated by large contributions due to opacity, 
indicates  that   there  is  tension  between  observational   data  and  SSMs
implementing OP  opacity tables.   Indeed, if we  restrict our analysis  to OP
opacities, i.e. we choose $\xi_{\rm opa}\equiv 0$ in our approach, the quality
of the  fit is considerably  decreased. The best  fit is obtained  for $\delta
z_{\rm CNO}  = 0.33\pm 0.03$  and $\delta  z_{\rm met} =  0.19 \pm 0.03  $ with
$\chi^2_{\rm min}/{\rm d.o.f.} = 66.9 / 32$ when all observational constraints
are  considered.   Solar  models  implementing OP  opacities  are  disfavoured
because they provide a less satisfactory  fit of the sound speed in the region
$0.3 < r/R_{\odot}  < 0.6 $, as it can be  seen from Figure~\ref{OPsound}.  It
is interesting to note that, when $\xi_{\rm opa}$ is allowed to vary, the best
fit  is   obtained  with  $\tilde{\xi}_{\rm  opa}\sim  1$   which  means  that
observational   data   are   better   described  when   using   OPAL   opacity
tables\footnote{We remind that we  used the fractional difference between OPAL
  and  OP  opacities  to   define  the  opacity  profile  uncertainty  $\delta
  \kappa_{\rm  opa}(r)$, see eq.~(\ref{dkappaOPALOP}).}.   The statistical
significance of this indication relies  on the correct evaluation of the sound
speed error in the outer radiative region of the Sun and may be weakened (or strengthened) 
by the possibility  of correlations in the inferences of the solar sound speed 
at different target radii (not considered here due to lack of the necessary information 
in the scientific literature).

\begin{figure}[t]
\par
\begin{center}
\includegraphics[height=4.6cm,angle=0]{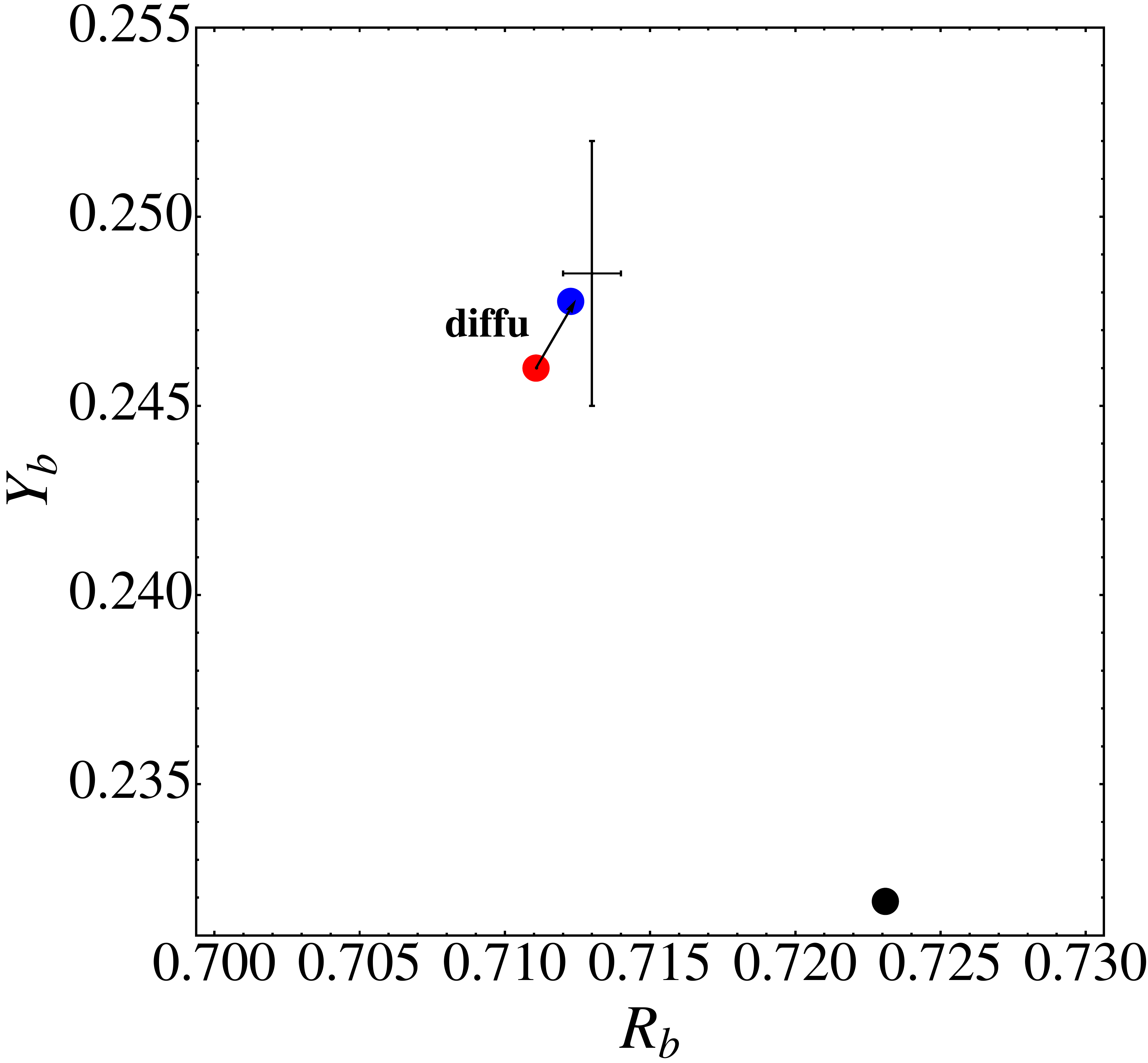}
\includegraphics[height=4.6cm,angle=0]{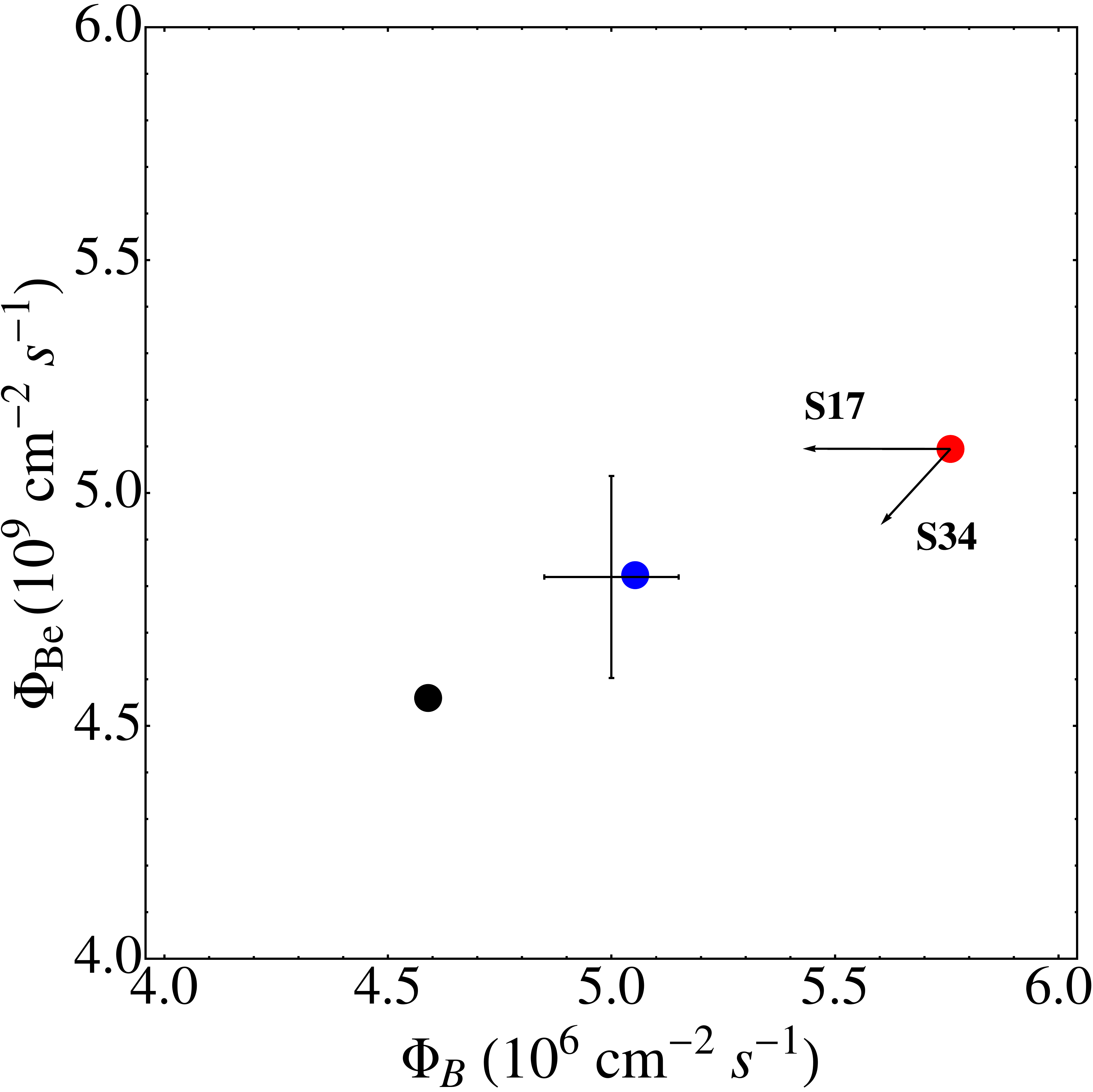}
\includegraphics[height=4.6cm,angle=0]{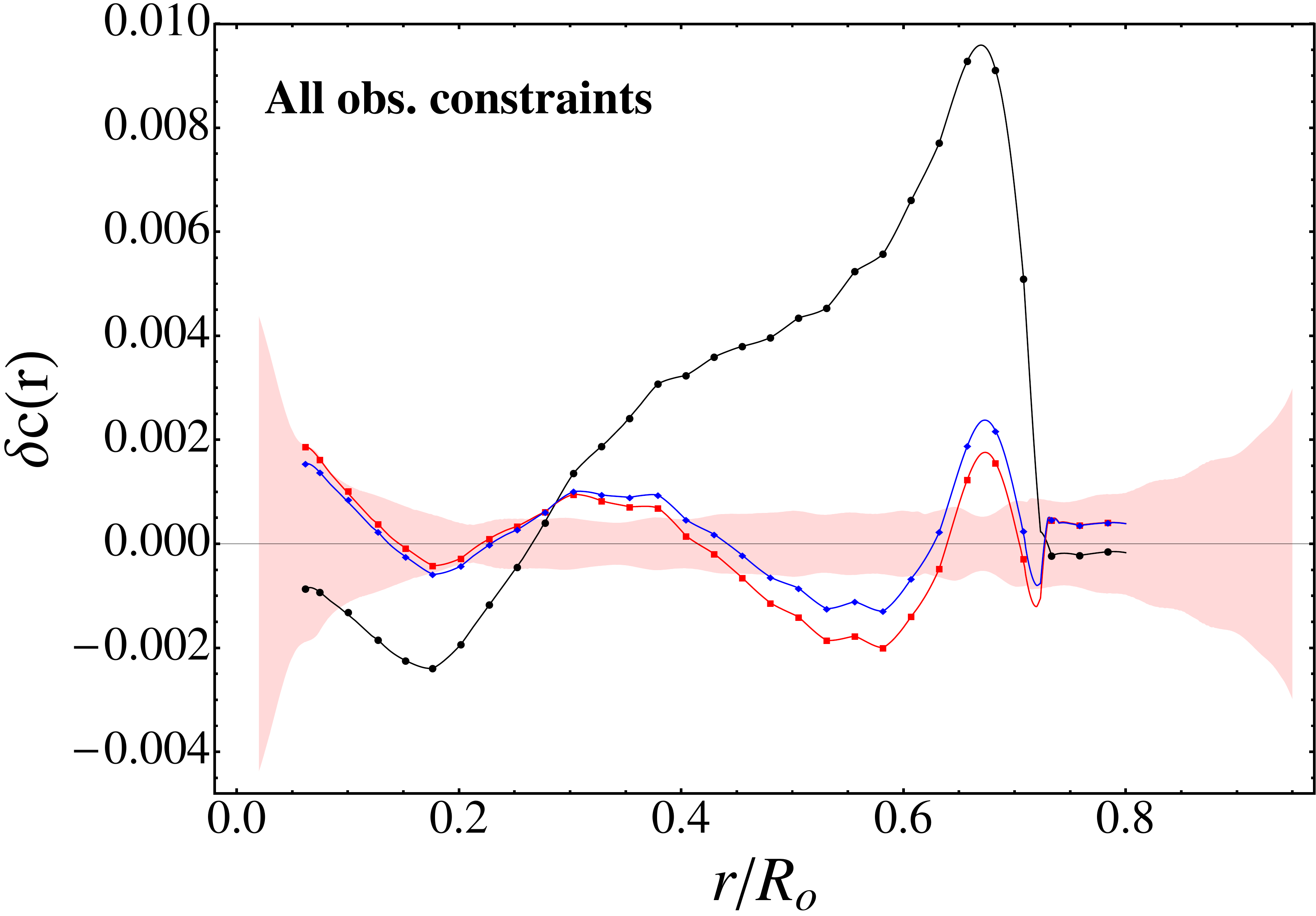}
\end{center}
\par
\vspace{-5mm} \caption{\protect Same as Figure~\ref{sound} but assuming
  $\xi_{\rm opa}\equiv 0$.
\label{OPsound}}
\vspace{0.5cm}
\end{figure}

%%--------------------------------
%\begin{figure}[t]
%\par
%\begin{center}
%\includegraphics[height=4.8cm,angle=0]{deropa.eps}
%\includegraphics[height=4.7cm,angle=0]{2PAR/All/gopaP.eps}
%\end{center}
%\par
%\vspace{-5mm}  \caption{\protect {\bf   Left  panel:}  The  logarithmic
% derivatives  of   opacity  with  respect  to   individual  metal  abundances
%  calculated along the  SSM profile. {\bf Right panel:}  The effective opacity
%  change  $\delta  \kappa(r)$ of  solar  models that  provide  a  good fit  to
%  observational constraints.}
%\vspace{0.5cm}
%\label{dkappa}
%\end{figure}
%%-----------

\item [5)]  The CNO  neutrino fluxes are  expected to  be $\sim 50\%$
  larger than those predicted by SSMs implementing AGSS09 composition.  Indeed,
  solar models providing a good  fit to the observational data give $\Phi_{\rm
    N}\simeq 3.4 \times 10^{8}\,{\rm  cm}^{-2}\,{\rm s}^{-1}$ and $\Phi_{\rm
    O} \simeq  2.5 \times 10^{8}\,{\rm  cm}^{-2}\,{\rm s}^{-1}$ as  a combined
  effect  of  the changes  in  composition  and, to  a  minor  extent, of  the
  systematic shifts in the input  parameters.  These values are even larger 
than predictions obtained by assuming GS98 surface composition.
 However, this result depends on
  the assumed  heavy element grouping.
% as it will be discussed in the next section.
 The  CNO neutrino fluxes,  in fact, are
  essentially determined by  the carbon abundance, see Table~\ref{LogDevComp},
  while  the  observational  data  included  in  our  analysis  are  basically
  sensitive to  the oxygen content of  the Sun, since this  element provides a
  large contribution to the solar opacity.
\end{enumerate}

\subsection{Three-parameter analysis}

\begin{figure}[t]
\par
\begin{center}
\includegraphics[width=5.4cm,angle=0]{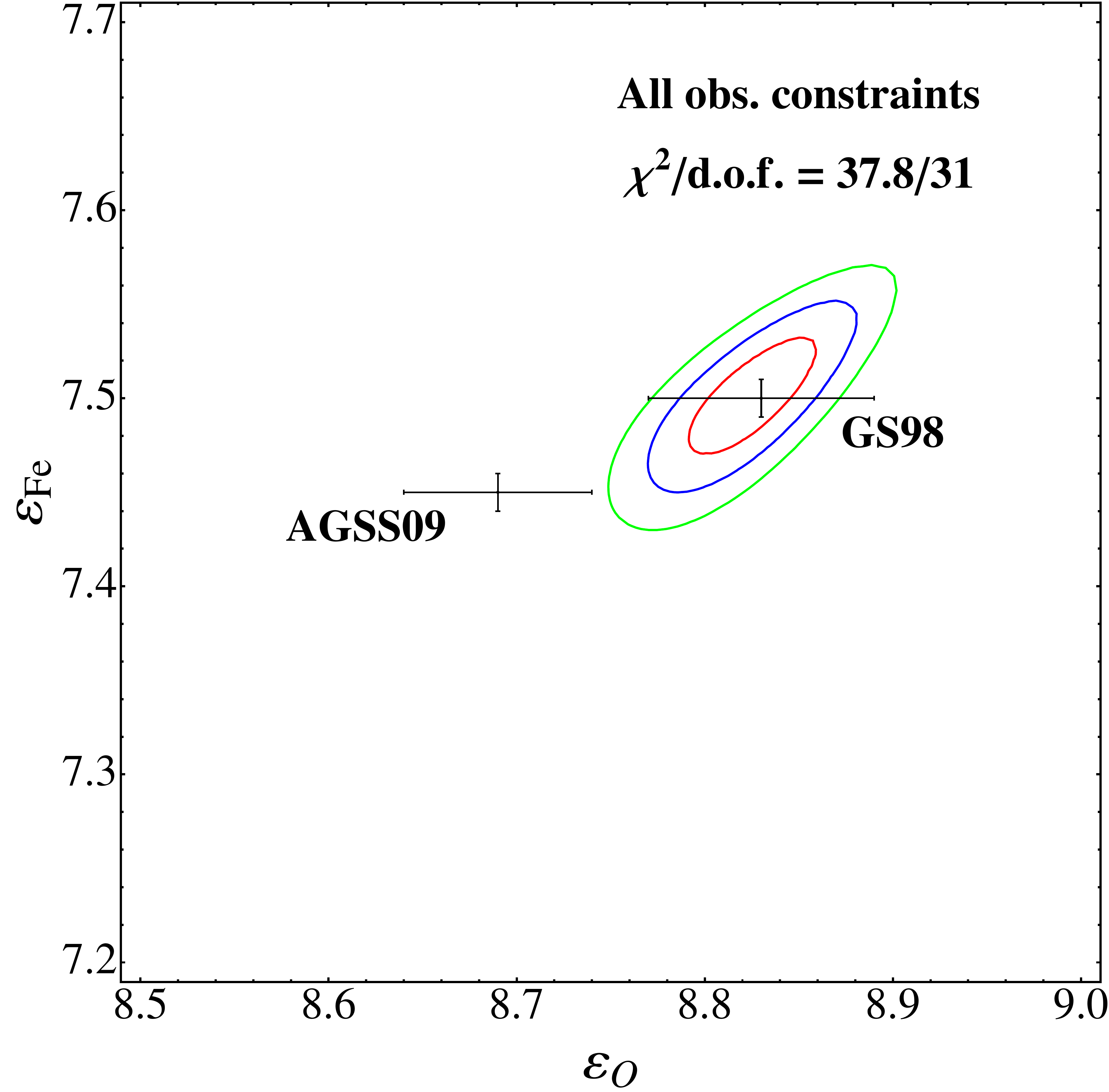}
\includegraphics[width=5.4cm,angle=0]{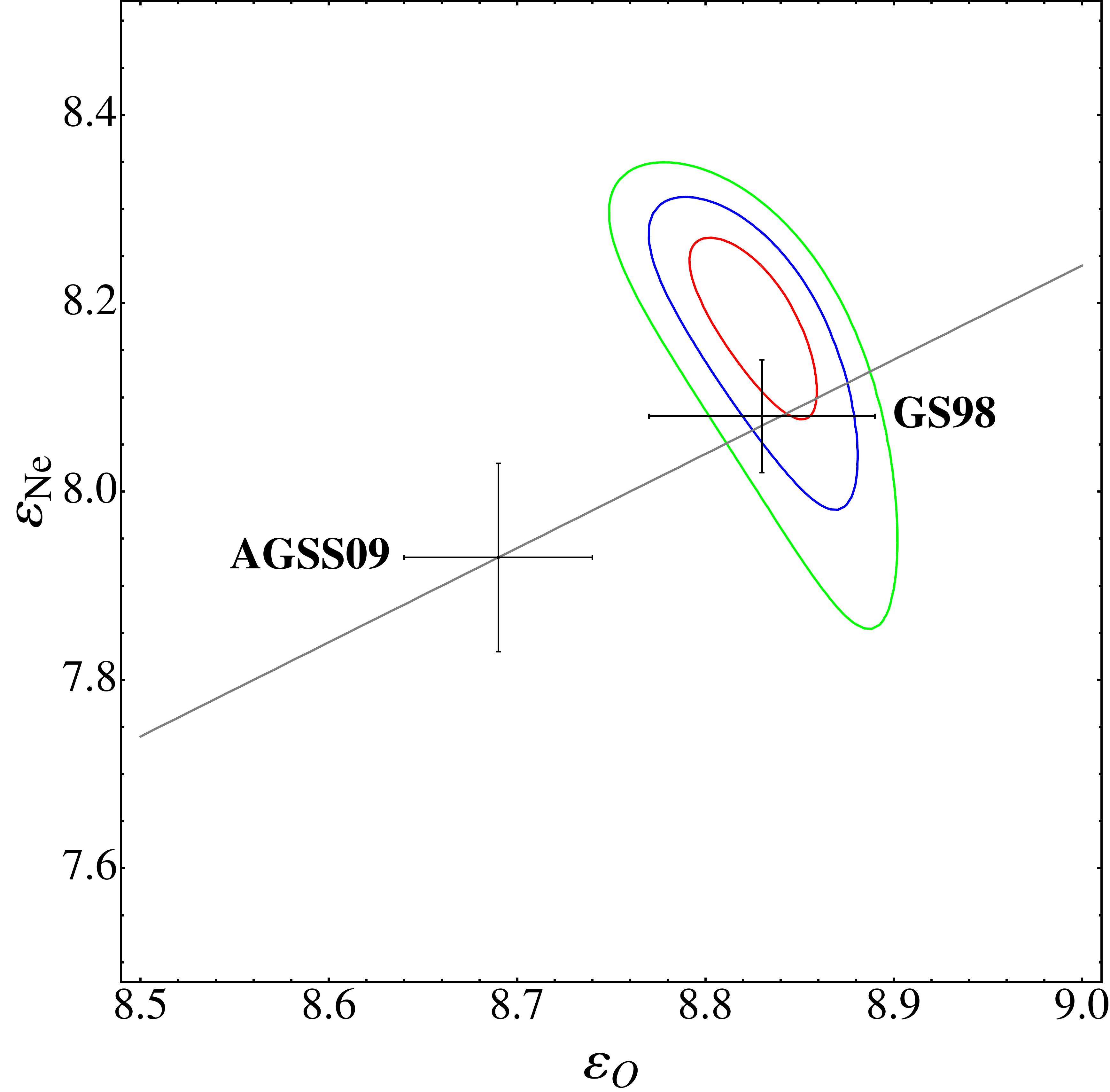}
\includegraphics[width=5.4cm,angle=0]{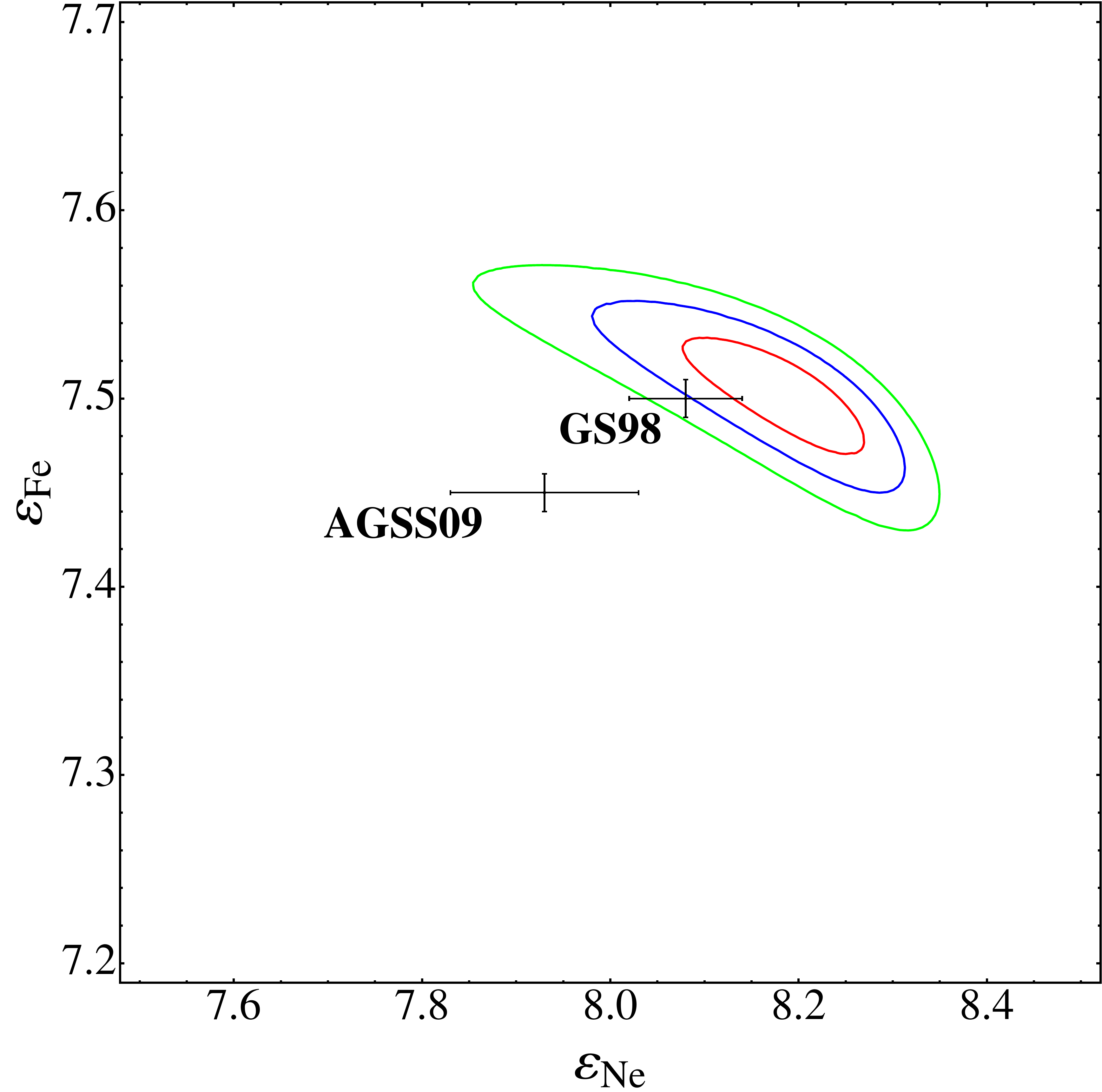}
\end{center}
\par
\vspace{-5mm} \caption{\protect The  bounds on $\varepsilon_{\rm O}$,
  $\varepsilon_{\rm  Ne}$  and $\varepsilon_{\rm  Fe}$  that  are obtained  by
  considering all  the available observational  constraints. The gray  line in
  the middle  panel corresponds to the  condition $\delta z_{\rm  Ne} = \delta
  z_{\rm  CNO}$,  i.e.  to  the  neon-to-oxygen  ratio  prescribed  by  AGSS09
  compilation.}
\vspace{0.5cm}
\label{results3par}
\end{figure}

It is important to discuss  how the above results changes when the
neon-to-oxygen  ratio  is allowed  to  vary,  since ${\rm Ne}$ lacks
photospheric features and the ${\rm Ne/O}$ ratio has to be inferred indirectly
from solar wind measurements.  In this  assumption,  the $\chi^2$  is
described as a function of  three parameters $\left(\delta z_{\rm CNO}, \delta
z_{\rm Ne}, \delta  z_{\rm met} \right)$ that can  be adjusted independently
to reproduce helioseismic and solar neutrino constraints.  In order to prevent
unphysical results, we add a penalty function to the $\chi^2$ given by:
\begin{equation}
\label{pen}
\chi^2_{\rm pen}  = \left[\frac{\delta z_{\rm Ne}-  \delta z_{\rm CNO}}{\Delta
    \, (1 + \delta z_{\rm CNO}) }\right]^2 
\end{equation}
where  $\Delta =  0.3$,  that forces  the  neon-to-oxygen ratio  to the  value
prescribed by AGSS09 compilation with a $1\sigma$ accuracy equal to $30\%$, as
it has been observed by \cite{neon}.
The bounds obtained by considering all the available observational constraints
are shown in Figure~\ref{results3par}.  The best fit composition is:
\begin{eqnarray}
\nonumber
\delta z_{\rm CNO} &=& 0.37 \pm 0.07 \\
\nonumber
\delta z_{\rm Ne} &=& 0.80 \pm 0.26 \\
\delta z_{\rm met} &=& 0.13 \pm 0.05
\label{bestfit3par}
\end{eqnarray}
that  correspond  to $\varepsilon_{\rm  O}=8.83  \pm 0.02$,  $\varepsilon_{\rm
  Ne}=8.19\pm 0.06$  and $\varepsilon_{\rm  Fe}=7.50 \pm 0.02$.   These values
are  still consistent  at $\sim  1  \sigma$ with  those obtained  in the  GS98
compilation. However,  the errors in  the inferred abundances are  larger than
before. We note, in particular, that  the neon abundance is bounded at the level
of  accuracy  prescribed  by  the  function~(\ref{pen})  indicating  that  the
observational data  are not effective  in constraining it.  The neon-to-oxygen
ratio is increased by about $\sim  30\%$ with respect to the AGSS09 value. The
quality of the fit, however,  is not significantly improved being $\chi^2_{\rm
  min}/{\rm d.o.f.} = 37.8/31$ and  the assumption $1+\delta z_{\rm Ne} = 1+\delta
z_{\rm CNO}$ is allowed at $1\sigma$.

The consequence of leaving neon as a free parameter is to introduce degeneracies between the 
various $\delta z_{j}$, as it is understood from inspection of Figure~\ref{results3par} and, in particular, 
by comparison of the left panel in Figure~\ref{results3par} and the lower-left panel 
of Figure~\ref{results2par}.
%As  it is  understood from  Figure~\ref{results3par},  the  consequence of
%leaving  neon as a  free parameter  is to  introduce degeneracies  between the
%various $\delta z_{j}$.   
It exists, in fact, a combination of
$\delta z_{\rm CNO}$, $\delta z_{\rm Ne}$  and $\delta z_{\rm met}$ 
that, taking also into account the effects of systematic pulls $\xi_{I}$, 
leaves substantially unchanged the observational properties of the Sun.
An increase of neon-to-oxygen ratio can be compensated by a
slight reduction of CNO and/or  Heavy elements according to simple approximate
formula:
\begin{eqnarray}
\nonumber
%\delta z_{\rm CNO} &=& 0.45 - 0.19 \, \left( \delta z_{\rm Ne} - \delta z_{\rm
%  CNO}\right)\\ 
\delta z_{\rm CNO} &=& 0.45 - 0.19 \, \Delta^{\rm Ne}_{\rm CNO} \\
%\left( \delta z_{\rm Ne} - \delta z_{\rm
%  CNO}\right)\\ 
\label{degeneracy}
%\delta z_{\rm  met} &=&  0.19 - 0.14  \, \left(  \delta z_{\rm Ne}  - \delta
%z_{\rm CNO}\right) 
\delta z_{\rm  met} &=&  0.19 - 0.14  \, \Delta^{\rm Ne}_{\rm CNO} 
\end{eqnarray}
where $\Delta^{\rm Ne}_{\rm CNO} = \delta z_{\rm Ne}  - \delta z_{\rm CNO}$, togheter with a re-adjustement of the systematic pulls:\footnote{
For simplicity, we report here only the systematic pulls that are sensitive the neon-to-oxygen ratio.
}
\begin{eqnarray}
\nonumber
\tilde{\xi}_{\rm opa} &=& 1.07 + 0.76 \, \Delta^{\rm Ne}_{\rm CNO}\\
\nonumber
\tilde{\xi}_{\rm diffu} &=& -0.74 - 0.41 \, \Delta^{\rm Ne}_{\rm CNO}\\
\nonumber
\tilde{\xi}_{\rm s_{33}} &=& 0.46 - 0.28 \, \Delta^{\rm Ne}_{\rm CNO}\\
\nonumber
\tilde{\xi}_{\rm s_{34}} &=& -0.97 + 0.58 \, \Delta^{\rm Ne}_{\rm CNO}\\
\tilde{\xi}_{\rm s_{17}} &=& -1.20 + 0.62 \, \Delta^{\rm Ne}_{\rm CNO}
\label{Pulls3Par}
\end{eqnarray}
From the above relations, we see  that a  $30\%$ uncertainty  in the  neon-to-oxygen ratio
roughly corresponds  to $\sim 6\%$ and  $\sim 4\%$ errors in  the inferred CNO
and heavy element abundances.  

This degeneracy can be discussed at a more basic level by  considering 
that the main effect produced by a change of the solar composition is the 
modification of the opacity profile of the Sun. The source term $\delta \kappa(r)$ that drives 
the modification of the solar  properties and  that is probed by  observational data can be  written as the
sum of two contributions \citep{villante2} :
\begin{equation}
\delta \kappa(r) = \delta \kappa_{\rm I}(r) + \delta \kappa_{\rm Z}(r)
\label{kappasource}
\end{equation}
The {\em intrinsic} opacity change, $\delta \kappa_{\rm I}(r)$, represents the
fractional variation of the opacity along  the SSM profile and it is given, in
our approach,  by $\delta \kappa_{\rm  I}(r) = \tilde{\xi}_{\rm  opa}\, \delta
\kappa_{\rm   opa}(r)$.   The   {\em  composition}   opacity   change  $\delta
\kappa_{\rm Z}(r)$ can be approximately calculated as:
\begin{equation}
\label{kappacomposition}
\delta    \kappa_{\rm    Z}(r)     \simeq    \sum_{j}    \frac{\partial    \ln
  \kappa(r)}{\partial \ln Z_{j}} \; \delta z_{\rm j}
\end{equation}  
by  using the  logarithmic derivatives  $\partial  \ln \kappa  / \partial  \ln
Z_{j}$  that are  presented in  the left  panel of  Figure~\ref{gOpa2}.  Taking
advantage  of  rel.~(\ref{kappasource}), we  calculate  the effective  opacity
change $\delta \kappa(r)$  that corresponds to the models  that provide a good
fit to observational data.  We see that $\delta \kappa(r)$ is well constrained
by the  available observational information.  
Opacity should be  increased by
$\sim \;  {\rm few}\%$ at  the center of  the Sun and by  $\sim 25 \%$  at the
bottom of the  convective envelope, as it was  calculated by \cite{villante2}.
The moderate increase at the  solar center improves the agreement with 
$Y_{\rm  b, obs}$ without  affecting the solar neutrino fluxes.  The increasing trend
of $\delta \kappa(r)$ is required to fit the convective radius $R_{\rm b}$ and
sound  speed profile  $\delta  c_{\rm i}$  (see  \cite{villante2}).  The  wavy
behaviour at intermediate radii improves consistency with inferred sound speed
values  in the  region $0.3  < r/R_{\odot}  < 0.6$.   The general  features of
$\delta \kappa(r)$  are essentially independent  on the assumptions  about the
opacity uncertainty. In  this respects, the increase of  the CNO and/or Ne content
is interpreted as providing the ``tilt'' to $\delta \kappa(r)$,
%as it is seen  from the shape of the various $\partial  \ln \kappa / \partial
%\ln Z_{j}$ in Fig.~\ref{.},
and is a solid conclusion of our analysis.

 Figure~\ref{gOpa2}, right panel,  also compares the effective variations of opacity  $\delta \kappa(r)$  
obtained in the two and three parameter analysis. In particular, the black dashed line 
corresponds to the  solar  model  with  the  composition given by equation
(\ref{bestfit3par}) and  the value $\tilde{\xi}_{\rm opa} = 1.40$ calculated from equation~(\ref{Pulls3Par}).  
The  red dotted line represents the effective opacity variation obtained with  parameters given by
equation~(\ref{bestfit2par}) and  $\tilde{\xi}_{\rm opa} = 1.07$.  
We see that the two lines coincide at the $2\%$ level or better. From this, we infer that
the reconstructed opacity profile does not depend on the assumed heavy element grouping. Moreover, 
we understand that the compositions (\ref{bestfit2par}) and (\ref{bestfit3par}) cannot  be discriminated 
by the  adopted observational  constraints.
More in general, they cannot be distinguished by any  conceivable observational test that 
is dominated by the opacity profile in the radiative region of the Sun:
the $2\%$ difference is indeed smaller than the   accuracy to which the opacity of the solar plasma 
is known. In summary, the neon-to-oxigen ratio cannot be effectively constrained with current data.

%=============================================================================

\section{Conclusions and perspectives}\label{sec:conclusions}

In this work, we have investigated the properties of the Sun by using a statistical approach,
normally adopted in other area of physics, in which the information provided by solar neutrino 
and helioseismic data can be combined in a quantitative and effective way.
Namely, we have inferred the chemical composition of the Sun by using the 
heliosesmic determinations of the surface helium abundance and of the depth of
the convective envelope; the measurements of $^7{\rm Be}$ and $^8{\rm B}$ neutrino fluxes; 
the solar sound speed profile inferred from helioseismic frequencies.

A consistent picture emerges from the combination of the different pieces of 
observational information which can be summarized as discussed in the following.\\
{\em i)} The surface composition prescribed by AGSS09 is excluded at an high confidence level, 
being the $\chi^2/{\rm d.o.f.} = 176.7/32 $ when all observational costraints are considered,
unless the SSM's chemical evolution paradigm is not correct and/or the opacity calculations are wrong; \\
{\em ii)}  A satisfactory fit to the available observational data ($\chi^2/{\rm d.o.f.} = 39.6/32$) 
is obtained in the context  of a two parameter analysis in which  volatile (i.e. C, N, O and Ne) 
and refractory elements (i.e. Mg, Si, S and Fe)  are grouped together and forced to vary by the 
same multiplicative factors. The abundance of volatile elements should be increased 
by $\left( 45\pm 4\right)\%$ while that of refractory elements should be increased by $\left( 19\pm 3\right)\%$ 
with respect  to AGSS09 values;\\
{\em iii)}  If the neon-to-oxygen ratio is allowed to vary within the currently 
allowed range (i.e. $\pm 30\%$ at $1\sigma$), the best fit compostion is obtained by 
increasing by $\left( 37\pm 7\right)\%$ the CNO elements; by $\left( 80\pm 26 \right)\%$
the neon; by  $\left( 13\pm 5\right)\%$ the refractory elements. The quality of the fit is, however, not significantly
improved with respect to the two parameter analysis, being $\chi^2/{\rm d.o.f.} = 37.8/31$.

 By taking advantage of the adopted statistical approach, we were  able to obtain few
additional conclusions concerning the properties of the Sun which are discussed in the following.\\ 
{\em iv)} Under the two and three parameter analyses, the CNO neutrino fluxes are expected to be 
substantially larger than those predicted by SSM implementing the AGSS09 surface composition, altough the exact value 
cannot be predicted in a model independent way since it depends on the assumed heavy elements 
grouping. In particular, this stems from assuming a the same fractional variation between C, N and O, a 
constraint that should be lifted when CNO neutrino fluxes are finally determined experimentally;\\
%{\em iv)}  The opacity profile of the solar radiative region is well constrained by the combination of 
%the different observational data.  The opacity profile of solar models that provide a good fit
%to observational constraints (rescaled to that of SSM implementing AGSS09 composition) 
%is shown by the gray band in the right panel of in Figure \ref{gOpa2};\\
{\em v)}  The sound speed in the region $0.3 < r/R_{\odot} < 0.6$  is better fitted  by using the old 
OPAL opacity tables rather  than the more recent OP opacity table. Indeed,  when we restrict
our analysis to OP opacities, the quality of the fit is considerably decreased giving 
$\chi^2/{\rm d.o.f.} = 66.9/32$; \\
{\em vi)} The observational data prefer values for the input parameters of the standard solar models 
that are slightly different from those presently adopted. Namely, the best fit is 
obtained by decreasing the diffusion coefficients by $\sim 10\%$ and the astrophysical factors 
$S_{34}$ and $S_{17}$ are decreased by $\sim 5\%$ and $\sim 9\%$ respectively, 
when all observational constraints are considered.
%{\em vi)} The CNO neutrino fluxes are expected to be substantially larger than those 
%predicted by SSM implementing the AGSS09 surface composition, altough the exact value 
%cannot be predicted in a model independent way since it depends on the assumed heavy elements 
%grouping.

\begin{figure}[t]
\par
\begin{center}
\includegraphics[width=8cm,angle=0]{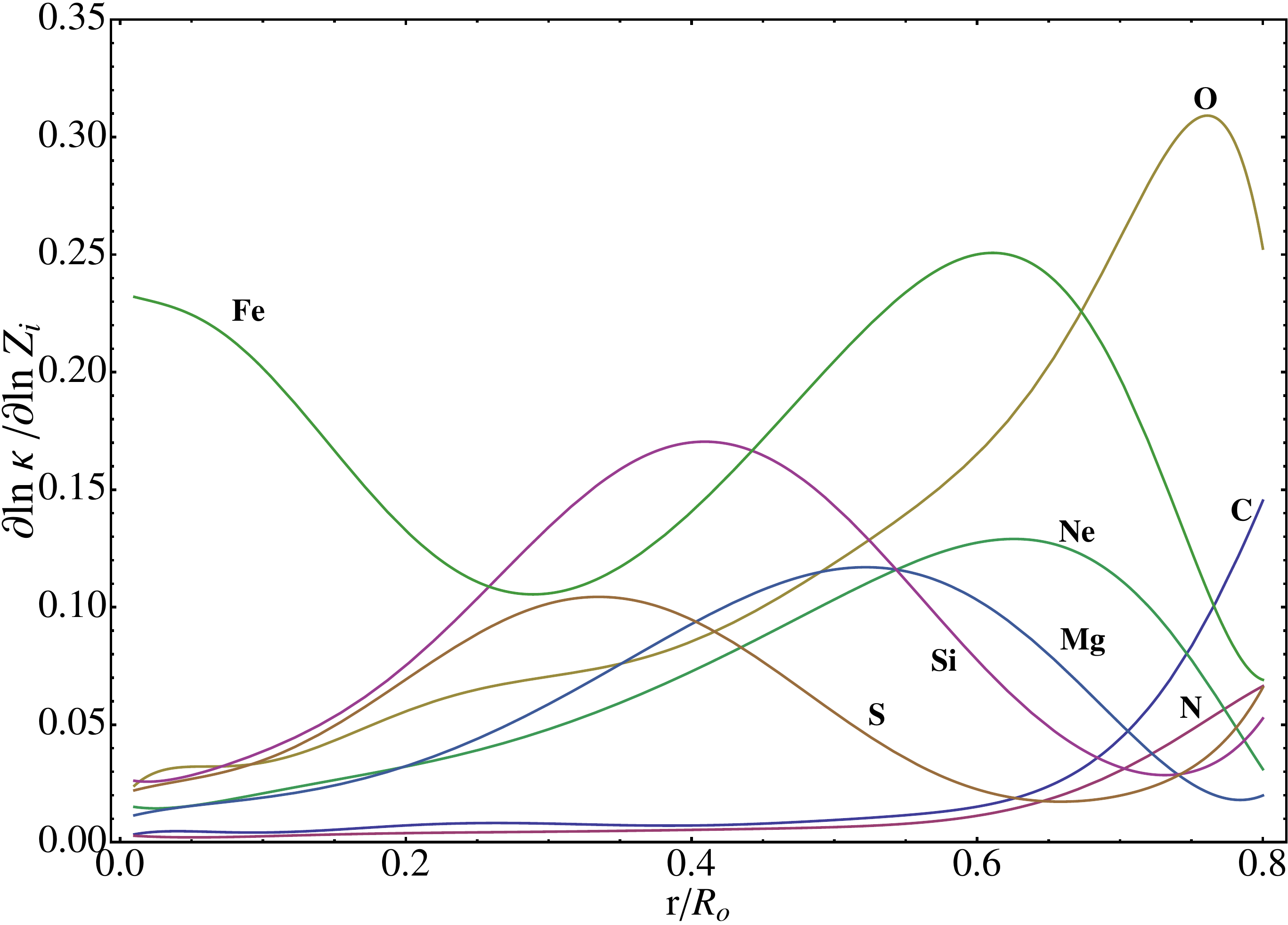}
\includegraphics[width=8cm,angle=0]{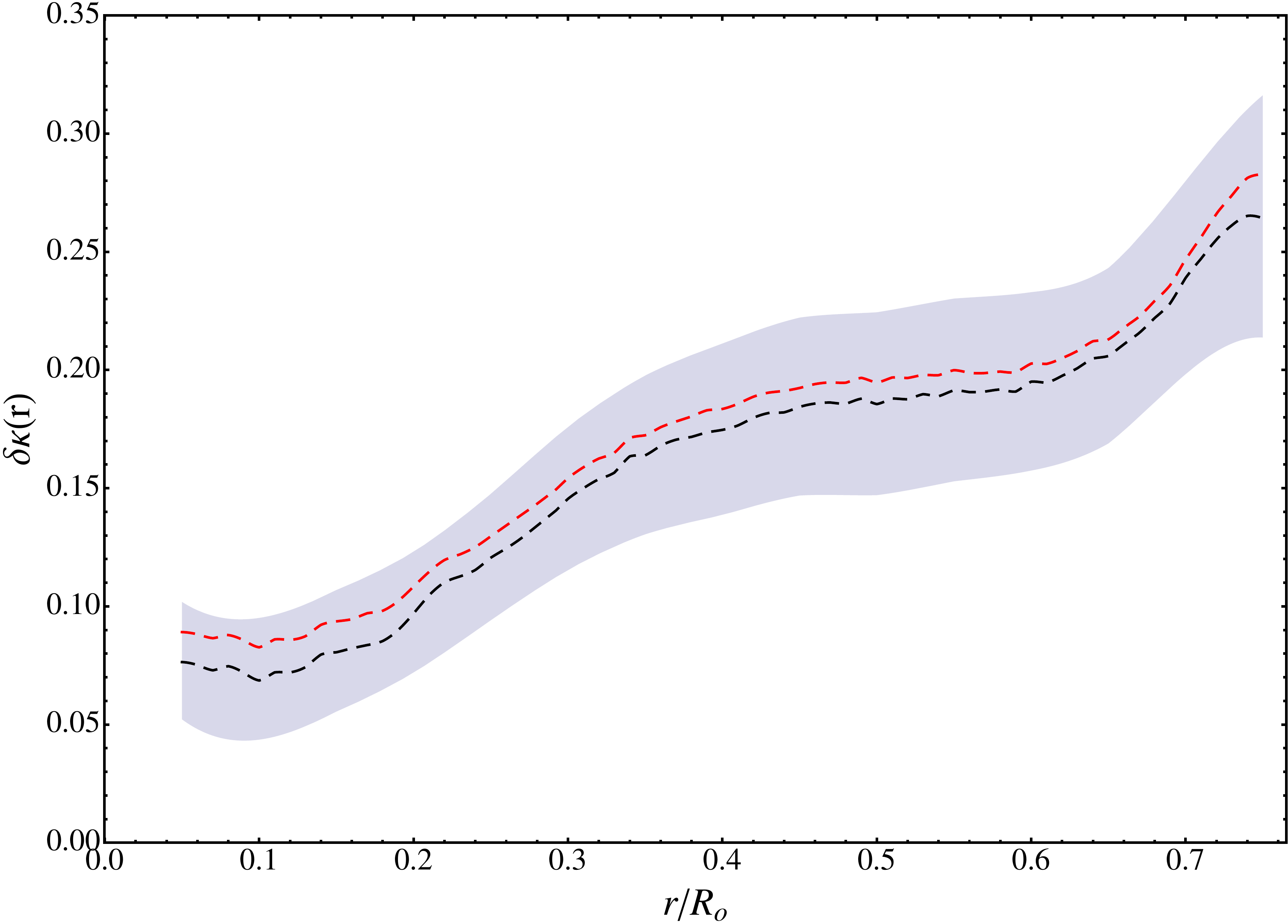}
\end{center}
\par
\vspace{-5mm} \caption{\protect 
Left  panel:  The  logarithmic derivatives  of   opacity  with  respect  to   individual  metal  abundances
calculated along the  SSM profile.  Right panel:
The  effective opacity change $\delta
  \kappa(r)$  of  solar  models  that  provide a  good  fit  to  observational
  constraints when $\left(\delta z_{\rm  CNO},\delta z_{\rm Ne}, \delta z_{\rm met}\right)$ are  allowed to vary.  The black dashed line  correspond to
  the best  fit model. The  red dashed line  correspond to the best  fit model
  obtained  with  the additional  assumption  that  $\delta Z_{\rm  Ne}=\delta
  z_{\rm  CNO}$, i.e.  that the  neon-to-oxygen ratio  is equal  to  the value
  prescribed by AGSS09 compilation.}
\vspace{0.5cm}
\label{gOpa2}
\end{figure}

The above results are obtained by using a simplified approach in which elements
are lumped togheter in two or three groups and they essentially follow from the fact that 
the opacity profile of the solar radiative region is well constrained by the combination 
of the different observational data, as it is shown by the gray band in the right panel of in Figure \ref{gOpa2}.
A substantial improvement with respect to the present situation could be provided   
by observational  constraints  where  the degeneracy  between  opacities and  composition  is  lifted.
One  such  constraint  has  already  explored  before and  is  linked  to  the
sensitivity  of  the  acoustic  p-modes  to  the  adiabatic  index  $\Gamma_1=
\frac{\partial  P}{\partial   \rho}_{\rm  ad}$.   Results   available  in  the
literature are contradictory.  \citet{Lin} concludes the metallicity of
the solar  envelope is comparable to  that of GS98.   However, using different
techniques  for  constructing the  solar  envelope  models  and the  inversion
procedures,  and also a  different equation  of state, \citet{Vorontsov}  
find a  solar metallicity that is even lower than  AGSS09 values. It is
important  to mention  that $\Gamma_1=  \frac{\partial  P}{\partial \rho}_{\rm
  ad}$,  while independent of  radiative opacities,  depends crucially  on the
details of the equation of state.
A  second possibility is  offered by  the neutrino  fluxes from  the CN-cycle.
While a  detailed quantitative analysis will be  presented elsewhere \citep{VillanteNext}, 
it is  important to stress at least qualitatively the
importance  of a  CNO neutrino  measurement. Even  a low  accuracy measurement,
providing a direct determination of the metallicity of the solar core, permits
to  remove the  degeneracy between  opacity  and composition  effects. Let  us
imagine e.g.  to measure the CNO  flux at the  20\% level. If the  detected fluxes
were  consistent with  the expectations  from  our analysis  (i.e. about  50\%
larger than  the reference predictions),  this would be sufficient  to conclude
that the  AGSS09 surface  abundances are wrong  and/or the  chemical evolution
paradigm of the  SSM is not correct. There would be  no possibility to explain
the  observed results by  assuming that  opacity (or,  more in  general energy
transport in  the Sun) is  not correctly described.   On the contrary,  if the
detected fluxes  were consistent  with those predicted  by solar models 
using AGSS09 admixture, then  this would  imply a  tension with  other observational  costraints. This
tension  could be  only explained  by assuming  that opacity  calculations are
wrong by a factor much larger than the presently estimated uncertainties. Both
these results would have enormous implications for stellar evolution.

\acknowledgements
A.S. is supported  by the MICINN grant AYA2011-24704 and by
the ESF EUROCORES Program EuroGENESIS (MICINN grant EUI2009-04170).

%---------------------------------------------------

\bibliographystyle{apj}

\end{document}